\documentclass[preprint,pre]{revtex4}
\usepackage{epsfig}
\usepackage{amsmath}
\begin{document}
\title{Transport and diffusion in the embedding map}
\author{N.Nirmal Thyagu}
\email{nirmal@physics.iitm.ac.in}
\author{Neelima Gupte}
\email{gupte@physics.iitm.ac.in}
\affiliation{Department of Physics, Indian Institute of Technology Madras, Chennai-600036,India}
\date{\today}
\begin{abstract}

We study the transport properties 
of passive inertial particles in a $2-d$ incompressible flows. Here, the particle dynamics  
is represented by the $4-d$ dissipative embedding map of the $2-d$ area-preserving
 standard map which models the incompressible flow. The system 
is a model for impurity dynamics in a fluid and is characterized by
two parameters, the inertia parameter $\alpha$,  and  the dissipation 
parameter $\gamma$. The aerosol regime, where the particles are denser than the fluid,
and the bubble regime where they are less dense than the fluid, correspond  
to the parameter regimes $\alpha>1$, and $\alpha<1$ respectively. 
Earlier studies of this system show a rich phase diagram 
with dynamical regimes corresponding to periodic orbits, chaotic structures and mixed regimes. 
We obtain the  statistical characterizers of transport for this system
in these dynamical regimes. These are the recurrence time statistics,
 the diffusion constant, and the distribution of jump lengths. The recurrence time 
 distribution shows a power law tail in the dynamical regimes where there is             
preferential concentration of particles in sticky regions of the phase space, and 
an exponential decay in mixing regimes. The diffusion constant 
shows behaviour of three types - normal, subdiffusive and superdiffusive, depending on the parameter regimes.  Phase
diagrams of the system are constructed to differentiate  different types of diffusion behaviour, as well as the behaviour of the absolute drift. 
We correlate the 
dynamical regimes seen for the system at different parameter values with
the transport properties observed at these regimes, and in the behaviour  of the transients. 
This system also shows the existence of a crisis and unstable dimension variability at certain parameter values. The signature of the unstable dimension
variability  is seen in the statistical characterizers of transport. We discuss the implications of our results
for realistic systems.

\end{abstract}
\maketitle

\section{Introduction}

The study of transport properties of impurities in fluid flows is a problem of  important  practical interest. These
properties can have serious implications for pollution in the atmosphere, plankton populations in 
the ocean, and diverse engineering applications. 
There is evidence that transport processes occurring in nature, e.g. the atmosphere \cite{newell}, 
are quasi-two dimensional \cite{haynes_book, pierre91}. The chaotic advection \cite{aref,texts_ottino} of impurities which are modelled by of inertial particles of finite size in flows of different types 
 has been studied both in the case of passive particles \cite{kuznetsov00,benczik03,vilela07r},
 and in the case of active particles that can react with the surroundings \cite{nishikawa02,reigada01}.  
Impurity dynamics in such two dimensional fluid flows can be effectively modelled by the 
 bail-out embedding maps of $2-d$ area-preserving maps \cite{motter03,nir07}.
We attempt to understand the dynamical and statistical properties of 
impurity transport in $2-d$ incompressible flows by studying such embedding maps.

The Lagrangian dynamics of small spherical tracers in  two dimensional
incompressible fluid flows is described by the Maxey-Riley equations.
These are further simplified under various approximations to give a
set of minimal equations called the embedding equations where the fluid
flow dynamics is embedded in a larger set of equations which include the
the differences between the particle and fluid
velocities \cite{motter03,cart02}.
Although the Lagrangian dynamics of the underlying fluid flow is
incompressible,
the particle motion is
compressible \cite{maxey87},
and has regions of contraction and
expansion. The density grows in the former giving rise to clusters and
falls in the latter giving rise to voids. The properties of the base
flow have important consequences for the transport and mixing of
particles. Map analogs of the embedding equations have also been constructed for cases where
the fluid dynamics is modelled by area-preserving maps which
essentially retain the qualitative features of the flow
\cite{pierre_fereday}.
The embedded  dynamics in both cases is dissipative in nature.
Here, we study the embedded standard map.

We study several  statistical characterizers of transport properties in different dynamical regimes of the system. These are the recurrence time statistics, diffusion and drift quantifiers, and the  distribution of jump lengths.  The first of these, viz. the  recurrence time statistics, show signatures of the dynamical regime of the system. The  cumulative probability distribution  of the recurrence  times (sometimes  called the complementary cumulative distribution \cite{altman06}), in the regime where chaotic structures are seen  shows the power law tail  characteristic of dynamics in an
inhomogeneous phase space with sticky regions. In contrast, in the regimes where full mixing is seen the RTD shows  exponential decays characteristic of rapid mixing. Earlier studies of the connection between the recurrence time statistics and the transport of particles had established  that normal transport occurs  
for cases where the recurrence times distribution shows  exponential behaviour, while 
anomalous transport occurs if there is a power law recurrence time distribution \cite{zaslav91,zaslav_physrep}. In the embedding map studied here, 
we observe that there  is a crossover from  exponential behaviour 
to power law behaviour in the recurrence time distribution in different dynamical regimes implying that there are some  dynamical regimes which are characterized by anomalous transport. The diffusion properties corresponding to these cases
are further explored. We also comment on the differences between the behaviour of aerosols and that of bubbles.

The  diffusion studies show that the system can show a variety of regimes corresponding to generalized 
diffusive behaviour. Three types of diffusion, viz. normal or brownian diffusion, superdiffusion, and subdiffusive behaviour are identified.  
A phase diagram is  constructed to 
classify and demarcate the regions of diffusion in the  $\alpha-\gamma$ parameter space.
Apart from the usually observed normal and superdiffiusive cases, the embedding map 
has two subclasses of subdiffusion -  one associated with trapping with stationary states and the other with trapping with non-stationary states.  The regions where ballistic diffusion is observed in the phase space are identified in the $\alpha-\gamma$ phase diagram.
Similarly, the areas where the average drift is zero are also likewise identified in the phase diagram. 
The diffusive behaviour of the system is compared with the dynamical regimes.
Our inferences can have implication for  the transport of pollutants in the atmosphere or other real application contexts.

The signature of the different dynamical regimes can also be seen in the transient
behaviour of the system. The jump length distributions, where the jump length is defined as the Euclidean distance between successive iterates, have envelopes which can be fitted by a heavy tailed Levy distribution in the chaotic structure regime which are characterized by sticky regions in the inhomogeneous phase space, and by a Gaussian envelope in the mixing regimes. 
The cumulated jump length distributions here show power-law behaviour for the sticky regimes, and cumulated Gaussians in the mixing regimes.  

In the aerosol regime of the embedding map, the plot of the largest
Lyapunov exponent shows signatures of unstable dimension 
variability (UDV) at certain parameter values. 
The recurrence time statistics shows the signature of the UDV, as does the diffusion constant. We discuss the consequences of this UDV for the transport properties of the system. We summarize the implications of our results.

\section{Dynamical regimes and phase diagram \label{dyn}}

The dynamics of inertial particles in a flow has been shown to be modeled by the bailout embedding equation \cite{motter03}

\begin{eqnarray}
\frac{d{\bf v}}{dt} - \alpha \frac{d{\bf u}}{dt}  & = & -\gamma({\bf v - u}).
\label{bailouteqn}
\end{eqnarray}
Here, the velocity of the particle is ${\bf v}=d{\bf x}/dt$ and the velocity field of the fluid flow is ${\bf u}(x,y,t)$, for a $2$-d fluid. 
 The  inertial parameter, $\alpha$  is related to the particle and fluid densities, $\rho_p$ and $\rho_f$ respectively,  by the equation $\alpha  =3 \rho_f/(\rho_f + 2 \rho_p) $. Thus, the $\alpha < 1$ regime corresponds to the aerosols, and the $\alpha > 1$ regime corresponds to the bubbles.  The dissipation parameter $\gamma$ is defined by  $\gamma = 2 \alpha/3 St$ where $St$ is the Stokes number,  and provides a measure of the contraction 
 or expansion in the phase space of the particle. The particle dynamics is dissipative in nature. 
A map analog of Eq. \ref{bailouteqn}  has been constructed \cite{cart02,motter03}. This  has the form

\begin{eqnarray}
{\bf x}_{n+1}  &=& {\bf M(x}_{n}) + \boldsymbol{\delta}_{n}\nonumber\\
\boldsymbol {\delta}_{n+1} &=& e^{-\gamma} [\alpha {\bf x}_{n+1} - {\bf M(x}_{n})].
\end{eqnarray}
Here, the area preserving map ${\bf M(x)}$ represents the incompressible fluid acting as the base flow. 
The vector $\bf{x}$ represents the position of the particle, and the vector $\boldsymbol{\delta}$ defines the detachment of the particle from the  fluid velocities  \cite{motter03}.  In our study, the dynamics of the base fluid flow has been represented  by the $2-d$ standard map, as it is a prototypical area preserving system \cite{chirikov79}, and  is  widely  used as a test bed  in a variety of  transport problems  \cite{white98}. The evolution equations  of the standard map are,
\begin{eqnarray} 
 x_{n+1}  &=&   x_n +  \frac{K}{2\pi}\sin( 2\pi y_n) , \nonumber\\
 y_{n+1}  &=&  y_n + x_{n+1}
\end{eqnarray} 
 
with  periodic boundary conditions such that $x,y \in [-1,1]$.   We fix the parameter $K=2$ in our study. The phase space of the standard map contains both regular and chaotic regions for this value.  Substituting the form of the standard map as ${\bf M}$ in the embedding map equation above, we get a $4$-dimensional embedding map,

\begin{eqnarray}
x_{n+1} & = & x_{n} + \frac{K}{2\pi}\sin(2\pi y_{n}) + \delta_{n}^{x}  \nonumber\\
y_{n+1} & = & x_{n} + y_{n} + \frac{K}{2\pi}\sin(2\pi y_{n}) + \delta_{n}^{y} \nonumber\\
\delta_{n+1}^{x} & = & e^{-\gamma}[\alpha x_{n+1}-(x_{n+1} - \delta_{n}^{x})] \nonumber\\
\delta_{n+1}^{y} & = & e^{-\gamma}[\alpha y_{n+1}-(y_{n+1} - 	\delta_{n}^{y})].
\label{bailstd}
\end{eqnarray}

\begin{figure}
\begin{center}
\resizebox{125mm}{95mm}{\includegraphics{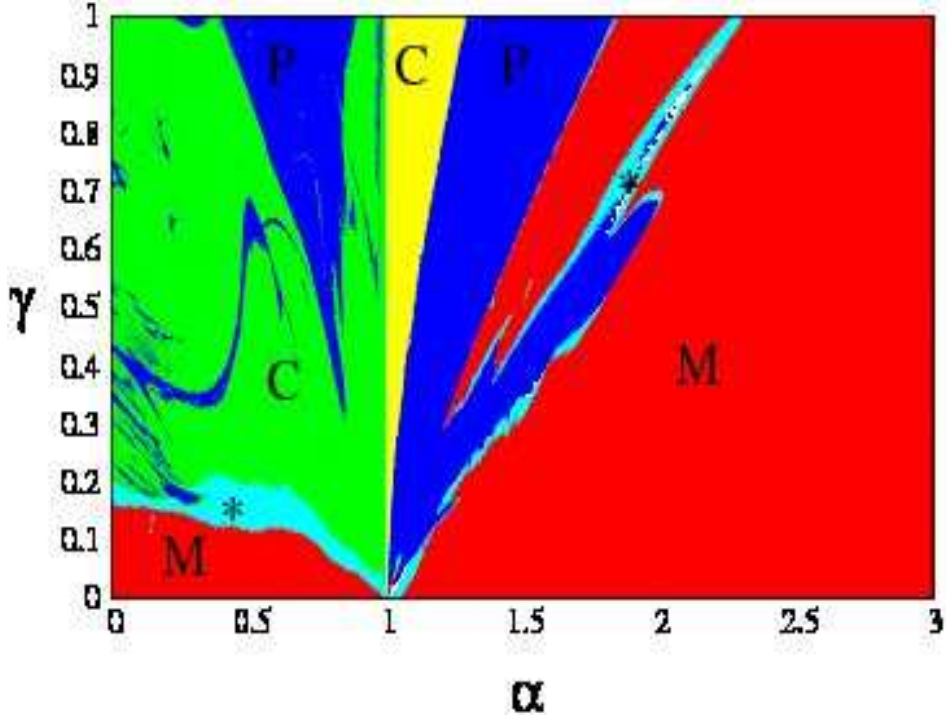}}\\

\end{center}
\caption{ (Color online) The phase diagram showing the three dynamical regimes in the embedding map in the $\alpha$-$\gamma$ parameter space. The regions marked by red (`M') show mixing and the regions marked by blue (`P') have periodic orbits. Chaotic structures are seen in the regions marked by green and yellow (`C'), in the aerosol and bubble regions respectively. There are some regions marked by light blue (labeled by `*') which have  chaotic structure over a mixing background in phase space\label{3dreg}}
\end{figure}

This map is invertible and dissipative.  We study the behaviour of the map in the parameter regimes
$0<\alpha< 3$, and $0 <\gamma < 1$.  Thus, our study encompasses both the aerosol regime i.e. 
$\alpha < 1$ and the bubble regime $\alpha >1$.  The dynamical regimes of the system have been catalogued in the phase diagram of Ref. \cite{nir07} for the same range of parameters.
A variety of dynamical behaviours can be seen at different values of $\alpha$ and $\gamma$.   These include periodic behaviour, chaotic structures
(Fig. \ref{sticky}(a)) and fully mixing regimes Fig. \ref{mix}(a). A detailed phase diagram of the system can be found in Fig.\ref{3dreg} ( See Ref.  \cite{nir07} for a detailed description of  the method of construction of the phase diagram) \footnote{In the context of the bailout embedding equation, Eq. \ref{bailouteqn},  the regime $\alpha > 3$ is unphysical as it implies negative particle densities.  While the $\alpha>3$ regime cannot be deemed unphysical by such a direct physical argument  in the context of the bailout  standard map, the regimes beyond $\alpha >3$ turn out to be mixing regimes for all values of $\gamma$, and hence have not been plotted in the phase diagram.}.

Since the map represents the Lagrangian dynamics of  particles, the chaotic structures seen due to the stickiness of the phase space correspond to the  preferential concentration of particles.
On the other hand, a uniform distribution of particles can be seen in the well mixed regimes of phase space. Thus, the phase diagram of the system has important implications for the transport properties of the system. In the next section, we examine  these transport properties via statistical characterizers.
We start with the recurrence time statistics.

\section{Recurrence Time Statistics \label{recur}}

The  statistics of recurrence times   are of fundamental importance in the study of chaotic systems \cite{gao,vbalki,vbalki2}.
These involve the study of recurrences of a given dynamical state of the system in finite time.
The first  recurrence time of a trajectory of a  system can be defined  to be the time $\tau_1$  taken for  a trajectory, which starts 
from a small subset $\xi$ of the phase space of the  system,  to return to the same subset $\xi$ in the limit where 
the volume of the subset $\xi \to 0$.  The trajectory may return to the subset at subsequent times 
$\tau_2,\tau_3,...\tau_n$ with  $\tau_n$ being called the $n^{th}$ recurrence time. A set  of the time intervals $\tau_i$ can be obtained in the  long time limit to give the recurrence time distribution of that trajectory.  The average recurrence 
time of the subset $\xi$ is calculated by averaging over the recurrence times of the trajectories starting in the subset $\xi$,  and the  average recurrence time  of the entire phase space   can be  obtained by averaging over the recurrence times of all the  partitions in the phase space. See Refs. \cite{zaslav_physrep,altmann05} for more rigorous definitions.

\begin{figure}[! t]
\resizebox{75mm}{75mm}{\includegraphics{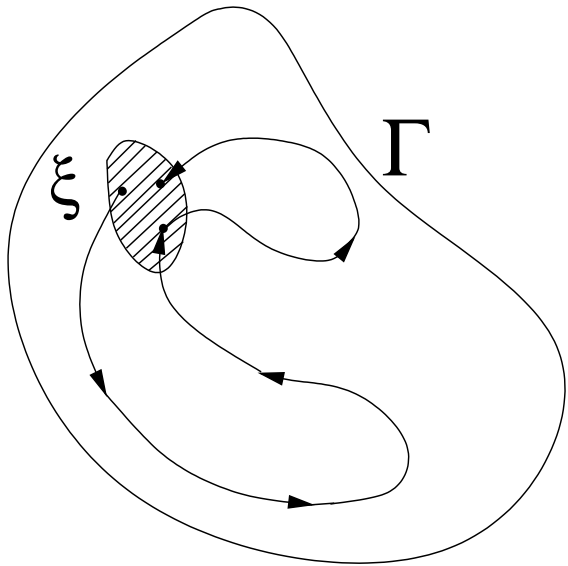}}\\
\caption{The schematic illustration of the recurrence phenomena in an invariant set $\Gamma$. A trajectory starting in a small partition $\xi$ is revisiting the partition in finite time. Here only the  first and the second recurrences are shown.\label{scheme}}
 \end{figure}

 Extensive studies of the mean recurrence
times have been carried out in the case of chaotic Hamiltonian systems \cite{chirikov99}.  
Generally, Hamiltonian systems are not fully hyperbolic, and the phase space of such Hamiltonian as well as that of area preserving systems contains 
regular islands and chaotic mixing regions in the phase space. A trajectory that originates in the chaotic region can stick to the neighborhood of the regular islands intermittently, and as a result, influence  long time properties like the recurrence times  giving rise to power law decays in the recurrence time distribution \cite{artuso08}.
The cumulative recurrence time distribution shows exponential decay  for uniformly hyperbolic system, where mixing is strong. In contrast, power law decay is observed for the recurrence time distributions of systems that have  inhomogeneous phase space \cite{altmann06}.

In the case of  area preserving maps, such as the standard map, the Poincar\'e recurrence time statistics shows two limiting cases. Firstly, for
the chaotic strong coupling limit (high values of the nonlinearity parameter $K$), where mixing is prominent in the phase space,  
the recurrence time distributions show  exponential decay as functions of time.
Secondly, for the near integrable weak coupling limit (small values of $K$, $K<K_{critical}  = 0.971 635 406 31$), the distributions show inverse power law behaviour \cite{buric}. At $K=K_{critical}$ the distribution shows a power law decay \cite{chirikov99}.

The base flow of our embedding map is the area-preserving  standard map, with the recurrence time statistics described above. However, the dynamics of the inertial particles, as described by the embedding map, is dissipative in nature.  It is therefore interesting to study the recurrence time statistics of the embedding map and see how the statistics of the inertial particles differs from the statistics of the base flow, and also see the effects of different dynamical regimes. We do this in the current section.
\begin{figure}[! t]
\begin{center}
\begin{tabular}{cc}
\hspace{0.2cm} (a)&
\resizebox{75mm}{75mm}{\includegraphics{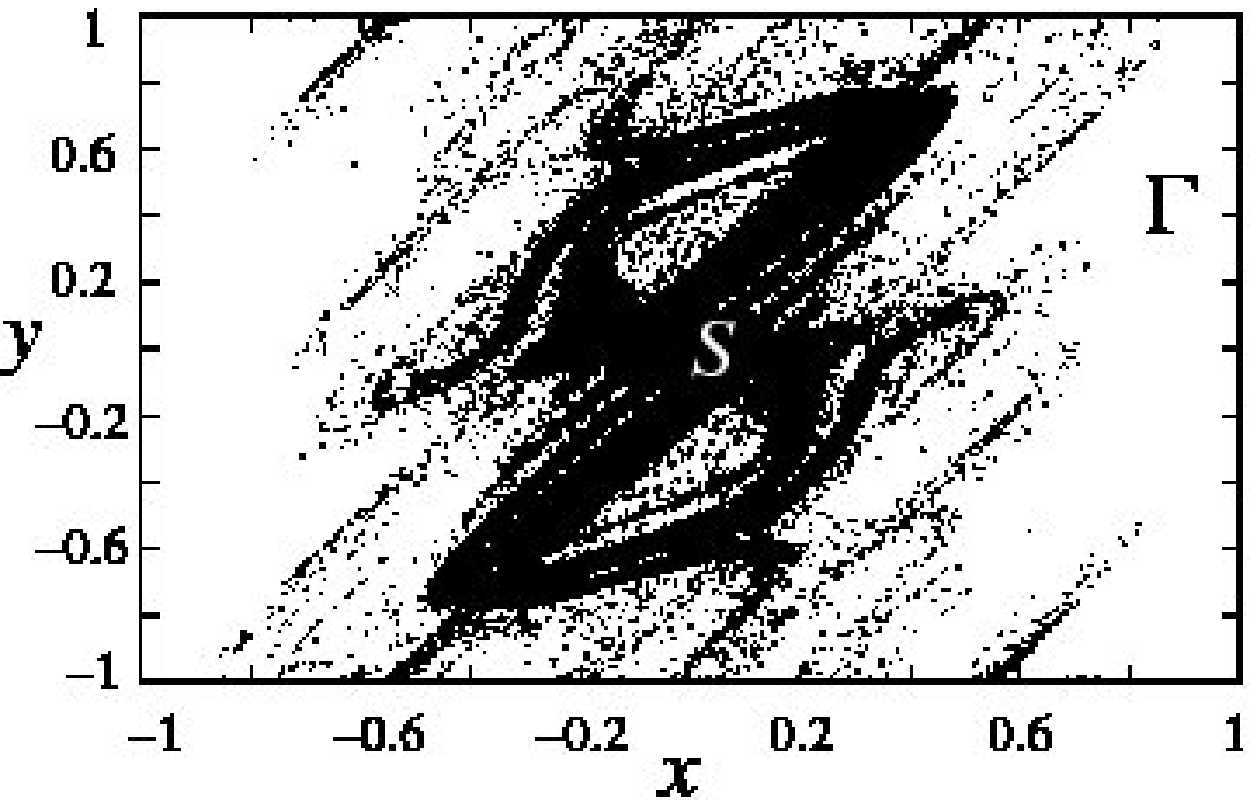}}\\
\end{tabular}
\end{center}
\begin{tabular}{cccc}
\hspace{0.2cm} (b)&
\resizebox{75mm}{75mm}{\includegraphics{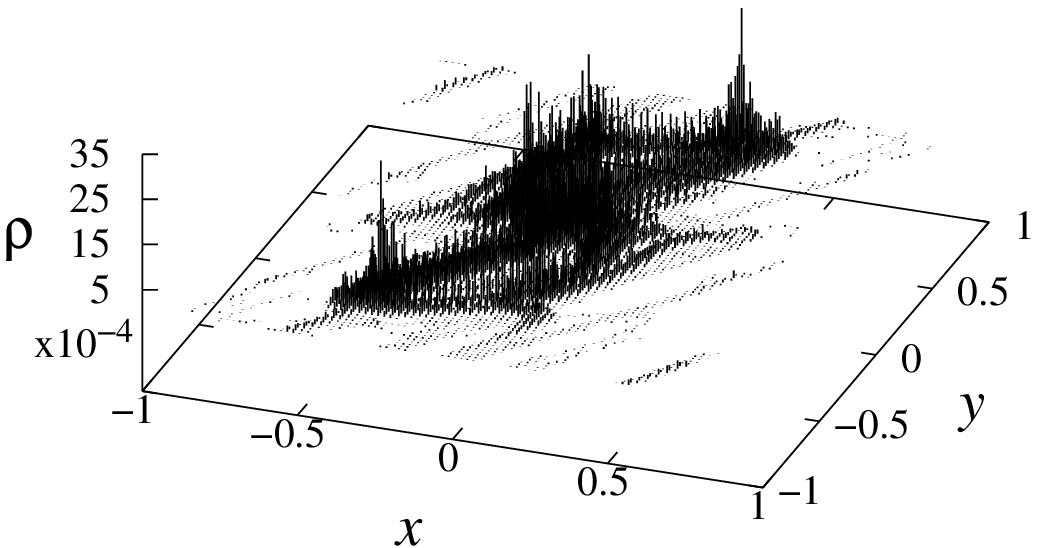}}&
\hspace{0.2cm} (c)&
\resizebox{75mm}{75mm}{\includegraphics{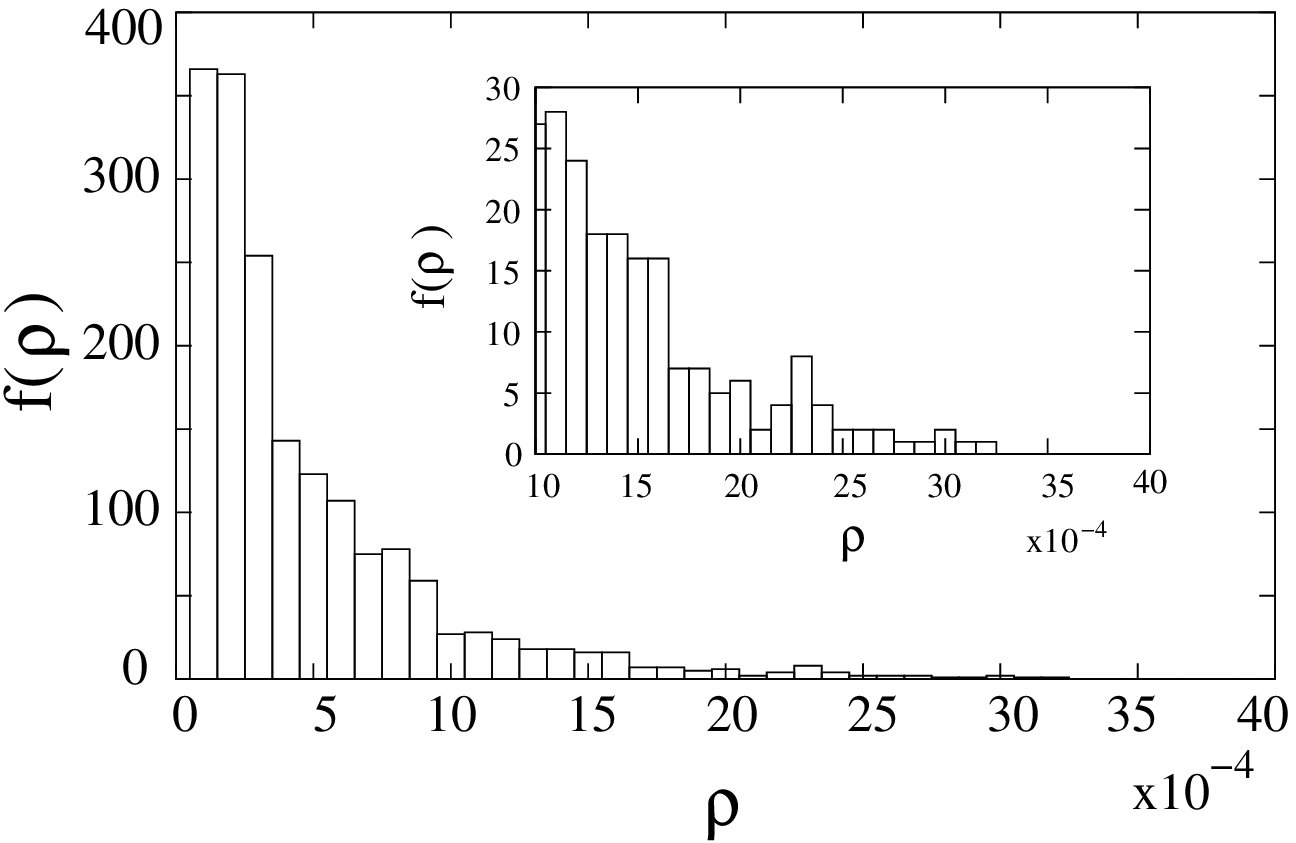}}\\
\end{tabular}
\caption{(a) The phase space plot with  chaotic structures ($\alpha=0.4$ and $\gamma = 0.4$). The darker regions are visited more often and the recurrence  times in these areas are longer. (b) The $3$-dimensional  invariant density plot of the attractor in the $x-y$ plane shows larger values of the invariant density $\rho$  where the region is sticky. (c) The histogram is plotted to show the distribution of the normalized invariant density $\rho$. \label{sticky}}
\end{figure}

We study the  recurrence time statistics  in the  $2$-dimensional phase space spanned by the configuration space coordinates $x-y$ and compare this with the recurrence time statistics in the $4$-dimensional phase space ($x,y,\delta_x,\delta_y$).  For  the $2-d$ statistics, the two dimensional configuration space $x,y$ with $x \in [-1,1]$ and $y \in [-1,1]$ is divided into a grid of $50$ x $50$ boxes and $200$  uniform random initial conditions are evolved for $1,000,000$ time steps, with $5,000$ iterations as transients. Each trajectory is associated with the box it visits immediately after the transients. The average recurrence time associated with any box is the recurrence time averaged over the initial conditions that are associated with that box \footnote{The effect of the finite size of the intervals in prototypical systems such as the logistic map and the Henon map, is well understood \cite{altmann04}.}.
      We compare the general features of the  recurrence time distributions for the two aperiodic dynamical regimes, viz., the chaotic structure regime and the mixing regime (see Fig. \ref{3dreg}) .

\begin{figure}[!t]
\begin{tabular}{cccc}
\hspace{0.2cm} (a)&
\resizebox{75mm}{75mm}{\includegraphics{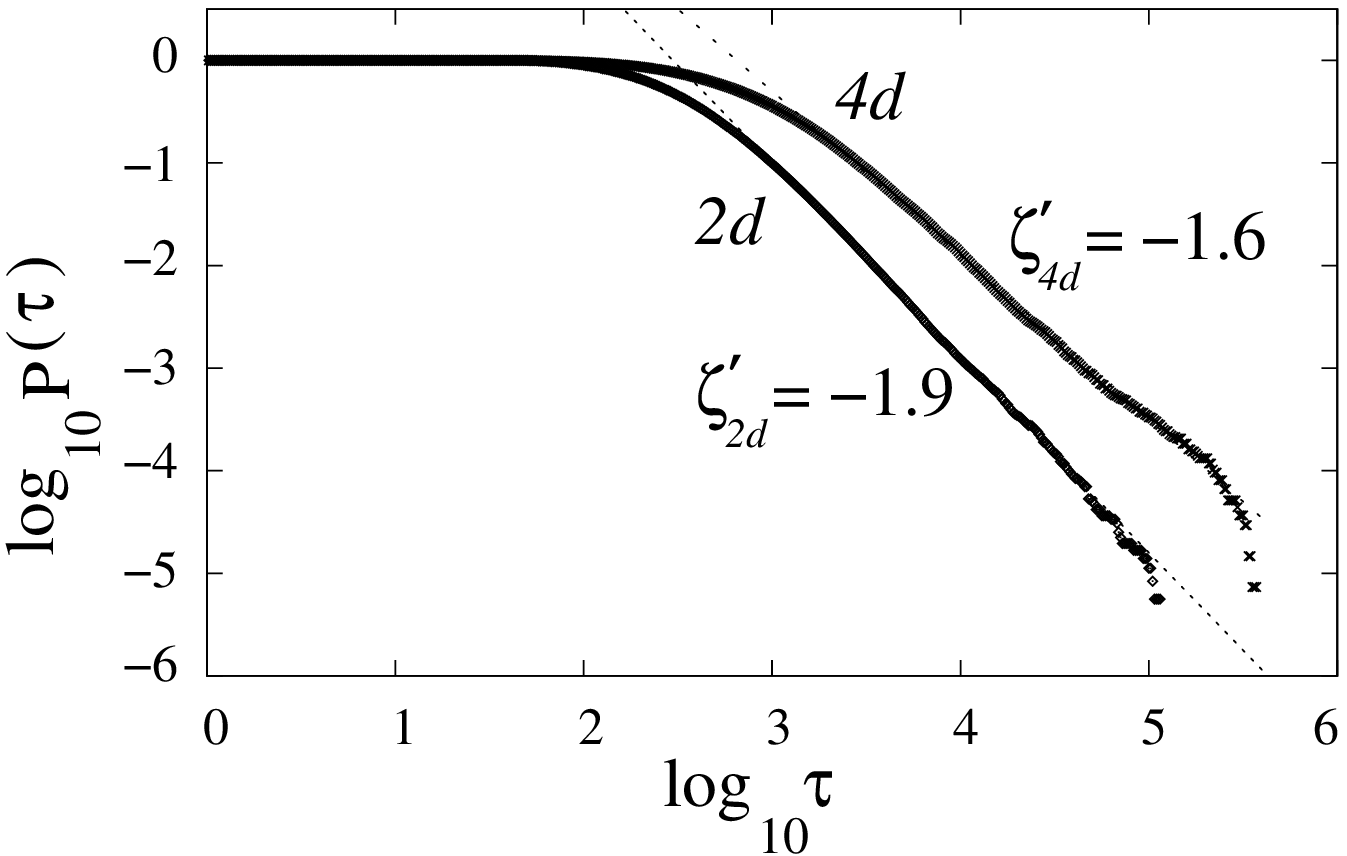}}&
\hspace{0.2cm} (b)&
\resizebox{75mm}{75mm}{\includegraphics{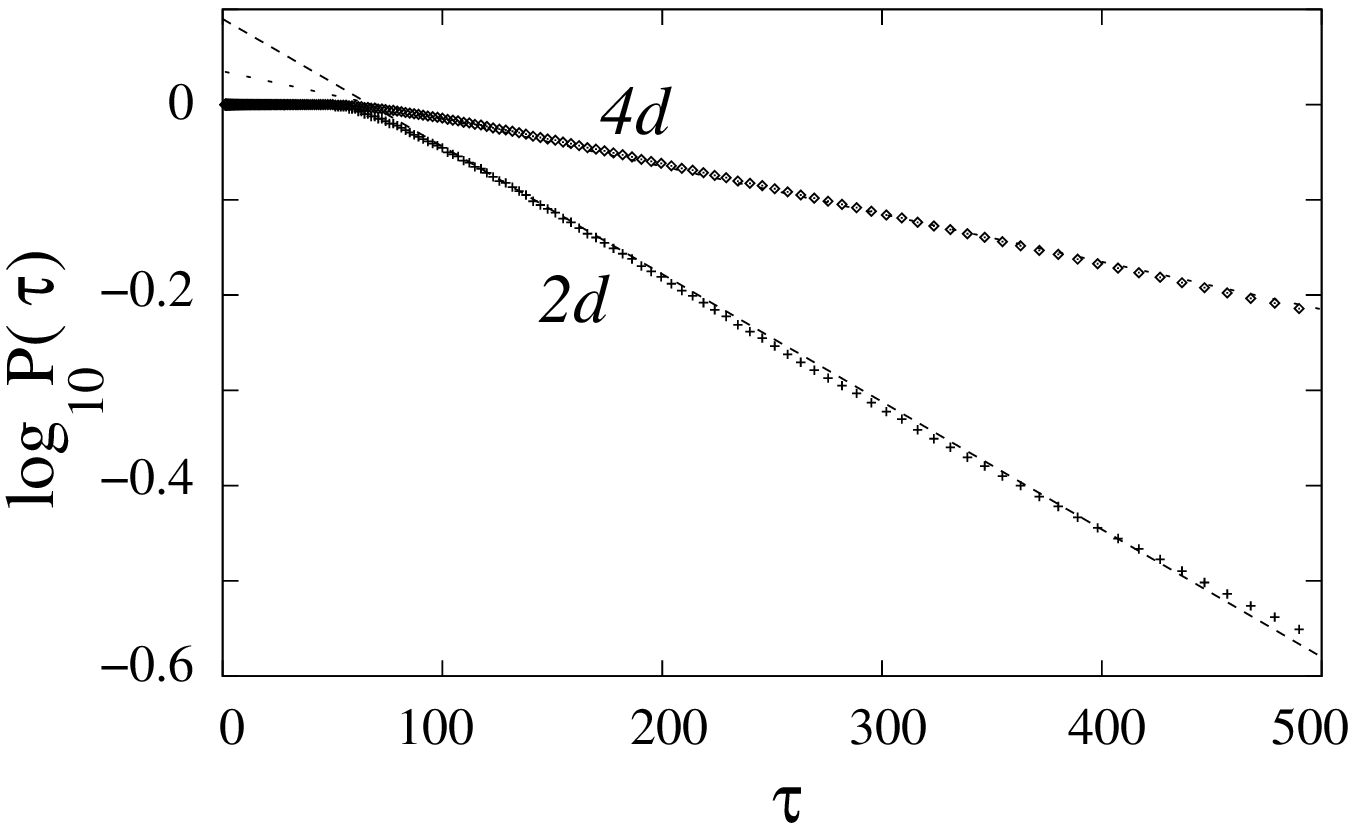}}\\
\end{tabular}

\caption{ (a) The complementary cumulative distribution of recurrence times at this parameter value shows a  power  law decay in the long time limit, with the exponent $\zeta_{2d} = \zeta'_{2d} -1 = - 2.9$ for the $2$-d  recurrence and $\zeta_{4d} = \zeta'_{4d}-1=-2.6$ for the $4$-recurrence. (b) The log-linear plot  in the short time limit shows an  an exponential decay, with the slope $ = -0.00134$ for the $2$-d case and slope $=-0.0005$ for the $4$-d case. \label{cstruct}}
\end{figure}

First, let us consider the chaotic structure regime  with the parameters $\alpha=0.4$ and $\gamma=0.4$. 
We first discuss the $2-d$ recurrences. It is clear from the $x-y$ phase space plot of Fig. \ref{sticky}(a) that the phase space in the chaotic structure regime is inhomogeneous. If the region available to all the trajectories for motion is called $\Gamma$,  there are some special regions $S$, contained in $\Gamma$ where the trajectory visits more often than others.  Hence, in  the phase space plots, the region  $S$  is  darker than the background $\Gamma$ (see Fig. \ref{sticky}(a)). These darker regions  are identified as the sticky regions, and have an important effect on the recurrence time statistics. The invariant density $\rho$ 
supported by the $x-y$ space is plotted in  Fig. \ref{sticky}(b). To  obtain the invariant density, random initial conditions were uniformly spread in the phase space covered by a grid, and the total number of times  that  trajectories visit a particular box in its itinerary was counted \cite{ott_book}.  It is clear that the value of the invariant density in the sticky regions is much higher than that in the background. 
The histogram of the normalized invariant density is plotted in Fig. \ref{sticky}(c) and clearly shows  that the major part of the phase space (the background) supports very small values of the invariant density, and the  large values of the invariant density are concentrated on a small subset  of the phase space, the sticky regions, or the regions where the chaotic structure is seen \footnote{The dynamic origin of the chaotic structures lies in an attractor widening crisis. The details of this crisis are discussed in \cite{nir07}.}.  Such sticky regions have  also been found in area preserving maps and experiments in fluids \cite{kuznetsov00,solomon}.

 The cumulative recurrence time statistics in the chaotic structure case  show a power law  decay for large times (see Fig. \ref{cstruct}(a)),
     
\begin{eqnarray}
P(\tau) \sim {\tau}^{ \zeta'}.
\end{eqnarray}

Here, the exponent of the cumulative probability distribution  $\zeta'$, and the exponent $\zeta$ of the probability distribution  are  related by $\zeta$ =$\zeta' -1$, as both the distributions show power law behaviour at large times.  In contrast, at short times, the cumulative RTD shows an exponential decay (see  Fig. \ref{cstruct}(b)), 
 
\begin{eqnarray}
P(\tau) \sim {e}^{ b \tau}.
\end{eqnarray}

We compare the $2-d$ recurrences seen here with the $4-d$ recurrences in the full $x-y$, $\delta_x-\delta_y$ space.  
The $4$-dimensional recurrence time distribution corresponding to the chaotic structure case is plotted along with the $2$-dimensional recurrence time distribution in Figs. \ref{cstruct}(a)-(b). 
Since the $2-d$ space is a projection of the $4-d$ space on the $x-y$ plane, 
recurrences to the $4-d$ space are expected to be longer than the recurrences to the $2-d$ space,
hence the distribution in $4-d$ has shifted to higher $\tau$ values. The character of the graph is unaltered, namely it shows a power law decay though the exponent is $\zeta_{4d}=-2.6$, while the  $2$-d exponent is $\zeta_{2d}=-2.9$. In the short time limit, where the distribution is exponential for both cases,  and the constant $b$  takes the value $b_{2d}=-0.00134$ and $b_{4d}=-0.0005$ for the $2$-dimensional and the $4$-dimensional recurrences respectively.  We note that the values of $\zeta$ and $b$ are not universal and change with the parameter  values of $\alpha$ and $\gamma$. However, the qualitative behaviour of the recurrence time distribution remains the same at other parameter values in the chaotic structure regime \footnote{
For values of the parameters $\alpha$ and $\gamma$  that show a power law decay at asymptotic times, the exponent is always found to be $\zeta<-2$. This is in conformity with the Kac's lemma  that for the distribution to have  finite moments the absolute value of the exponent should be greater than $2$ \cite{zaslav_physrep,zaslav_text}.}.

In the case of the area preserving maps, the power law scaling indicates that  the phase space is inhomogeneous. In the case of the dissipative map discussed here, a similar effect is seen due to the existence of the chaotic structure or sticky regions in the phase space. 
The examination of the residence times of  trajectories  starting from arbitrary initial conditions, shows that the trajectories spend $80 \%$ of their time in the sticky regions, regardless of the initial conditions. Trajectories which start in the non-sticky regions of the phase space
wander into the sticky regions and spend a long time there before emerging outside. This leads to long recurrence times. Other trajectories have short recurrence times. 
The exponential decay seen in the early part of the RTD corresponds  to the short recurrences.
On the other hand, the  power law scaling regime is due to the recurrences to the non-sticky regions. This accounts for the crossover from  exponential to  power law behaviour  seen in the recurrence time distribution. Thus, the chaotic structures play the same role in the phase space for the embedding map as the islands seen in the area-preserving maps. However, there is no hierarchy in the chaotic structures, unlike that in the case of the hierarchical islands of the area-preserving maps.
\begin{figure}[! t]
\begin{tabular}{cccc}
\hspace{-0.8cm} (a)&
\resizebox{75mm}{75mm}{\includegraphics{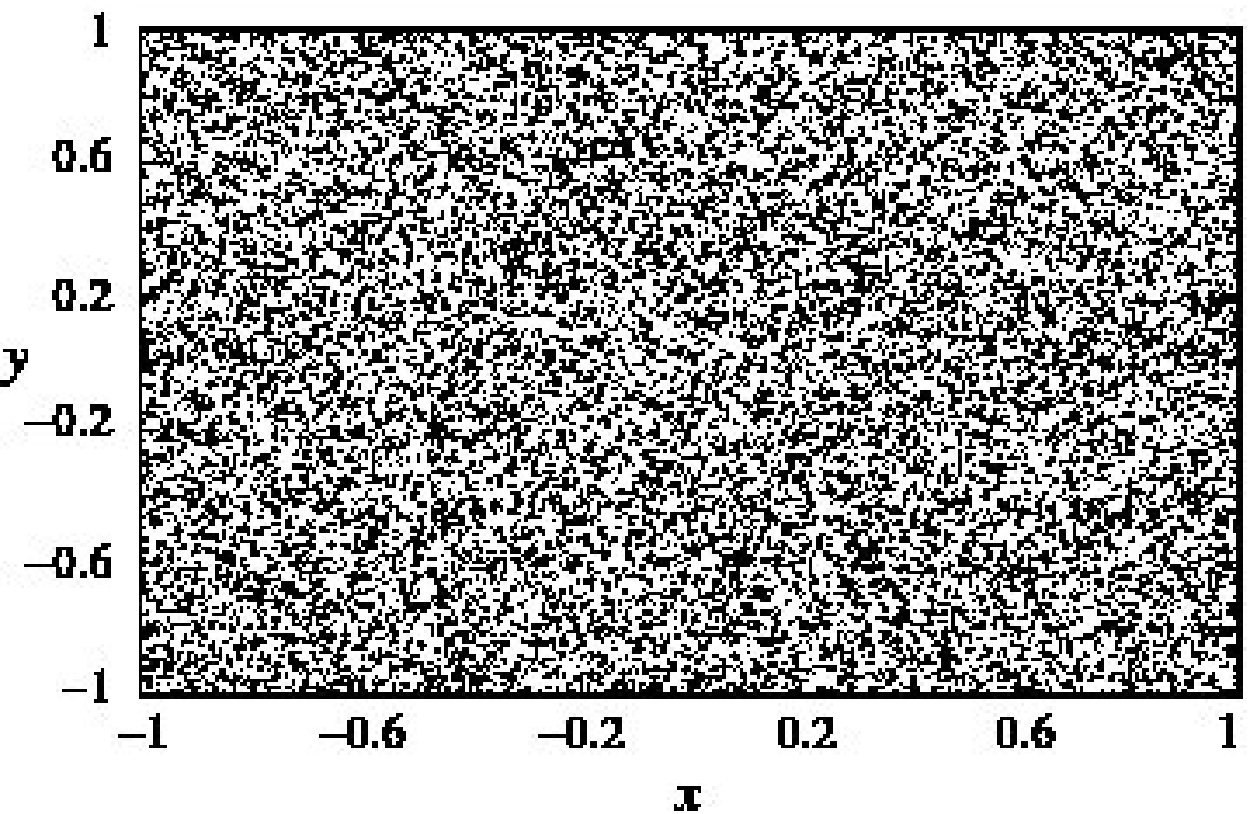}}&
\hspace{0.8cm} (b)&
\resizebox{75mm}{75mm}{\includegraphics{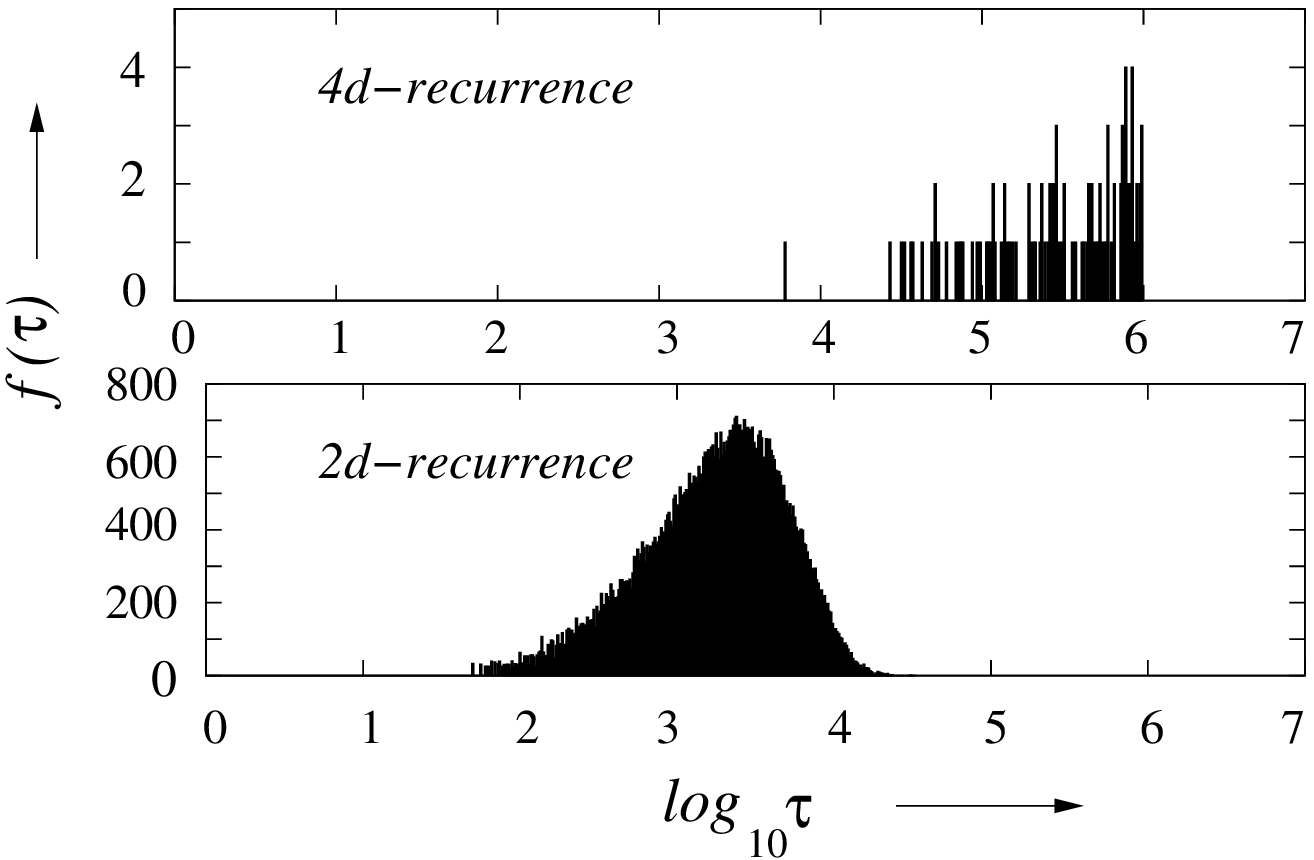}}\\

\hspace{-0.8cm} (c)&
\resizebox{75mm}{75mm}{\includegraphics{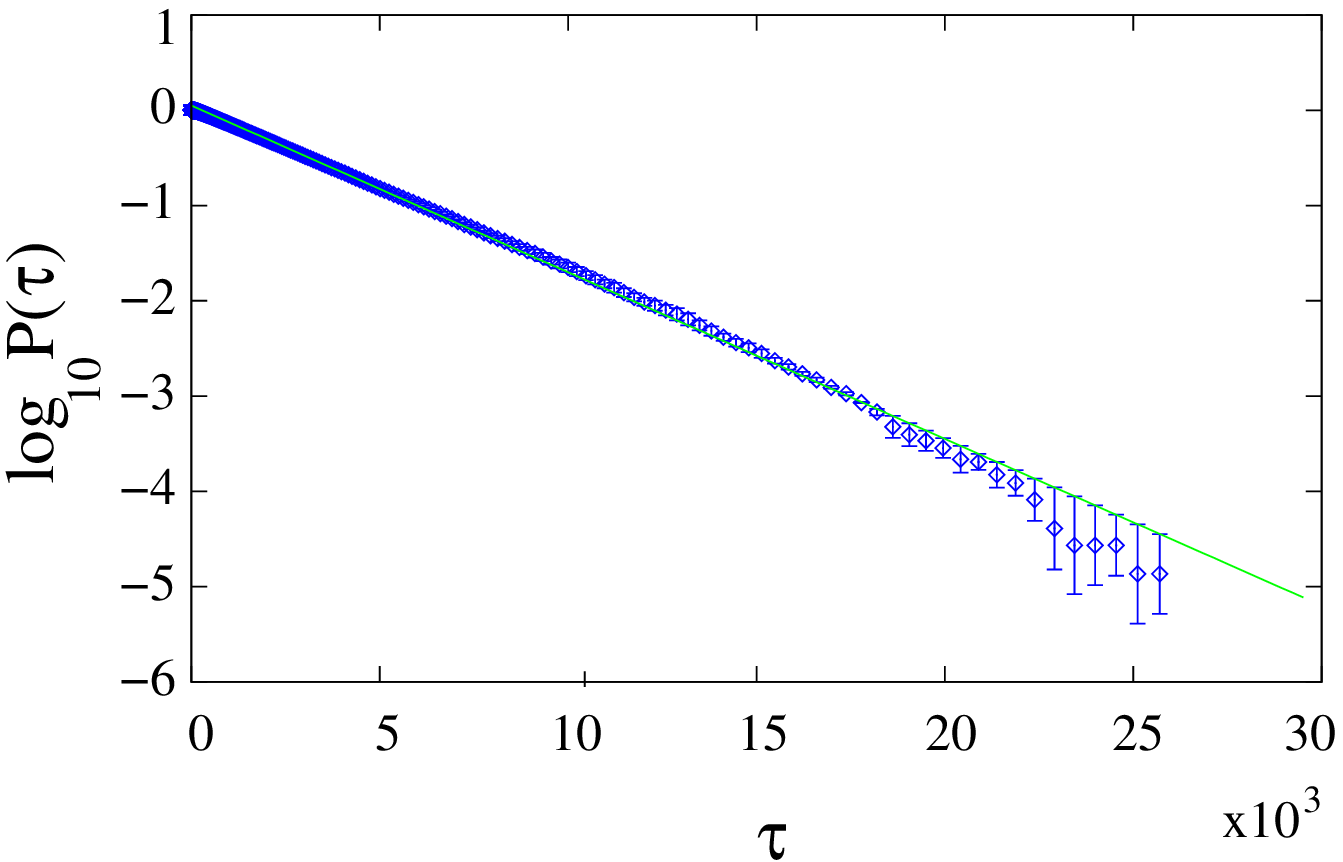}}&
\hspace{0.8cm} (d)&
\resizebox{75mm}{75mm}{\includegraphics{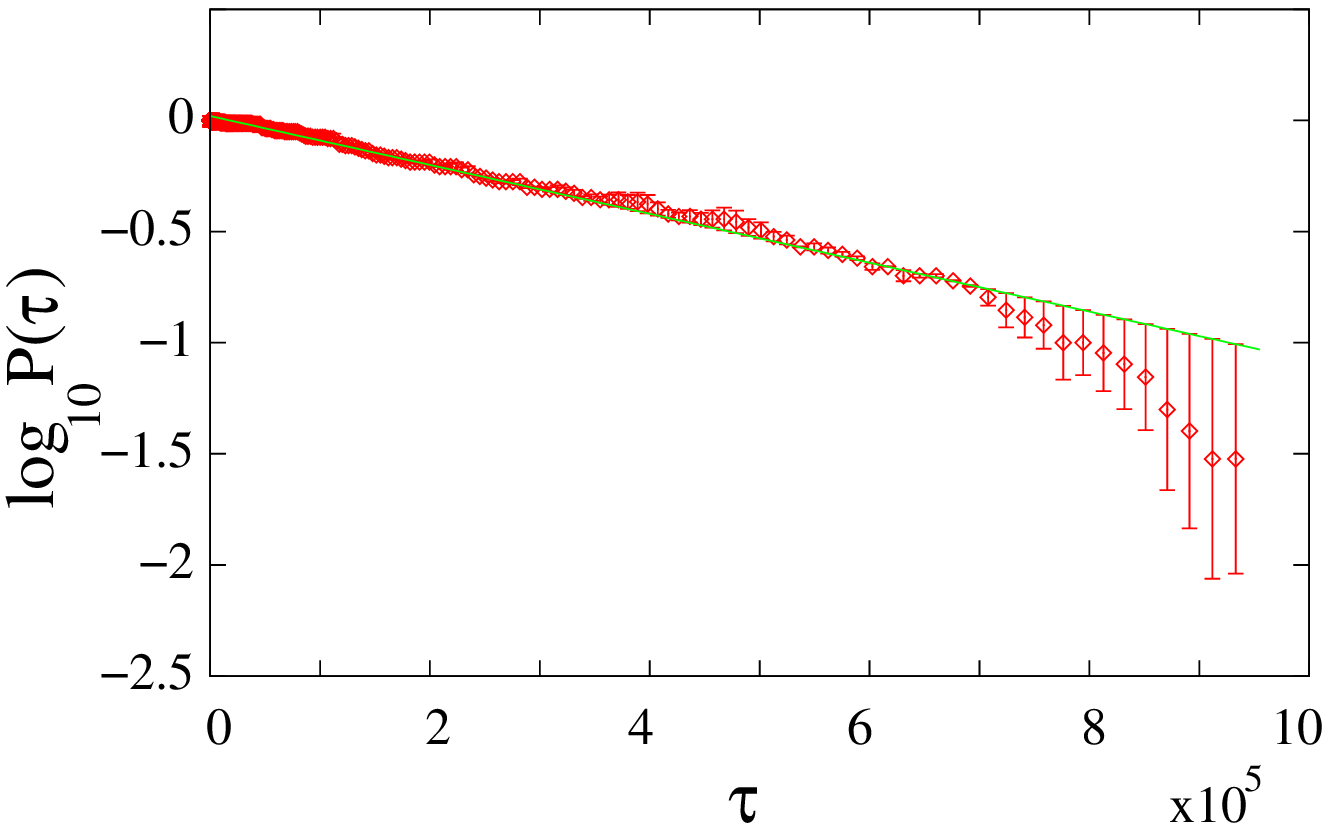}}\\
\end{tabular}

\caption{(a) The phase space plot in a mixing regime ($\alpha=2.8$ and $\gamma = 0.4$). We can see there is no preferential concentration of particles in any region of the phase space. (b) The histograms of the distribution of recurrence times in  $4$-dimensional case (top) and $2$-dimensional case (bottom).  The log-linear plots of the complementary cumulative  distribution of  recurrence times in (c) $2$-dimensions with slope $-0.000175$, and (d) $4$-dimensions with slope $-0.0000011$. \label{mix}}
\end{figure}
    
The behaviour of the recurrence time distribution in the  mixing regimes
of the phase diagram is quite different. Here, unlike the chaotic structure case, the mixing is very strong and there are no preferential regions in the phase space which the  particles visit more frequently, or stick to for longer times.   The phase space plot of a typical mixing regime can be seen in Fig. \ref{mix}(a). The recurrence time distributions show exponential decay both in the $2$-d and the $4$-d case, as is expected for the strongly mixing case. In Figs. \ref{mix} (c) and (d) the log-linear plots of the RTD shows the exponential decays with the slopes for the $2$-dimensional and $4$-dimensional cases to be $-0.000175$ and $-0.0000011$ respectively. The histograms of the recurrence times for the $2$-dimensional and the $4$-dimensional cases are shown in Fig. \ref{mix}(b).  It is clear that the  $4$-dimensional recurrence times are longer than the $2$-dimensional recurrence times. Also it is  evident that the frequency of occurrence of  the $4$-d recurrences is much smaller  than the frequency of occurrence of the  $2$-d recurrences (compare  the  values of $f(\tau)$ in both the ordinates of Fig. \ref{mix}(b)).

In the chaotic structure case we saw a crossover from the exponential decay behaviour at short times to power law behaviour at asymptotic times. This crossover behaviour seen  in the  recurrence  statistics signals a transition in the transport properties of the system \cite{zaslav91}.   We examine the transport properties using detailed diffusion studies,  and  correlate them  with the recurrence times and the dynamical regimes.

\section{Diffusion \label{diffusion}}
The transport of passive inertial particles in flows can be described statistically by
examining the dispersion as a function of parameters $\alpha$ and $\gamma$.  We consider an ensemble of N initial particles distributed uniformly randomly in the phase space and evolve them in time. As the initial conditions evolve in time, the particle cloud drifts in the two dimensional configurational space from the initial position  as well as the individual particles disperse from the  moving cloud. The dispersion of particles is
given by the variance of the displacement of 
particles $\sigma^2$,

\begin{eqnarray}
\sigma^2 (t) = <({\bf x}(t)-<{\bf x}(t)>)^2 > \sim D t^\eta
\label{eqn-diff}
\end{eqnarray}

\begin{figure}[! t]

\begin{tabular}{cccc}
\hspace{-0.8cm}(a)&
\resizebox{75mm}{75mm}{\includegraphics{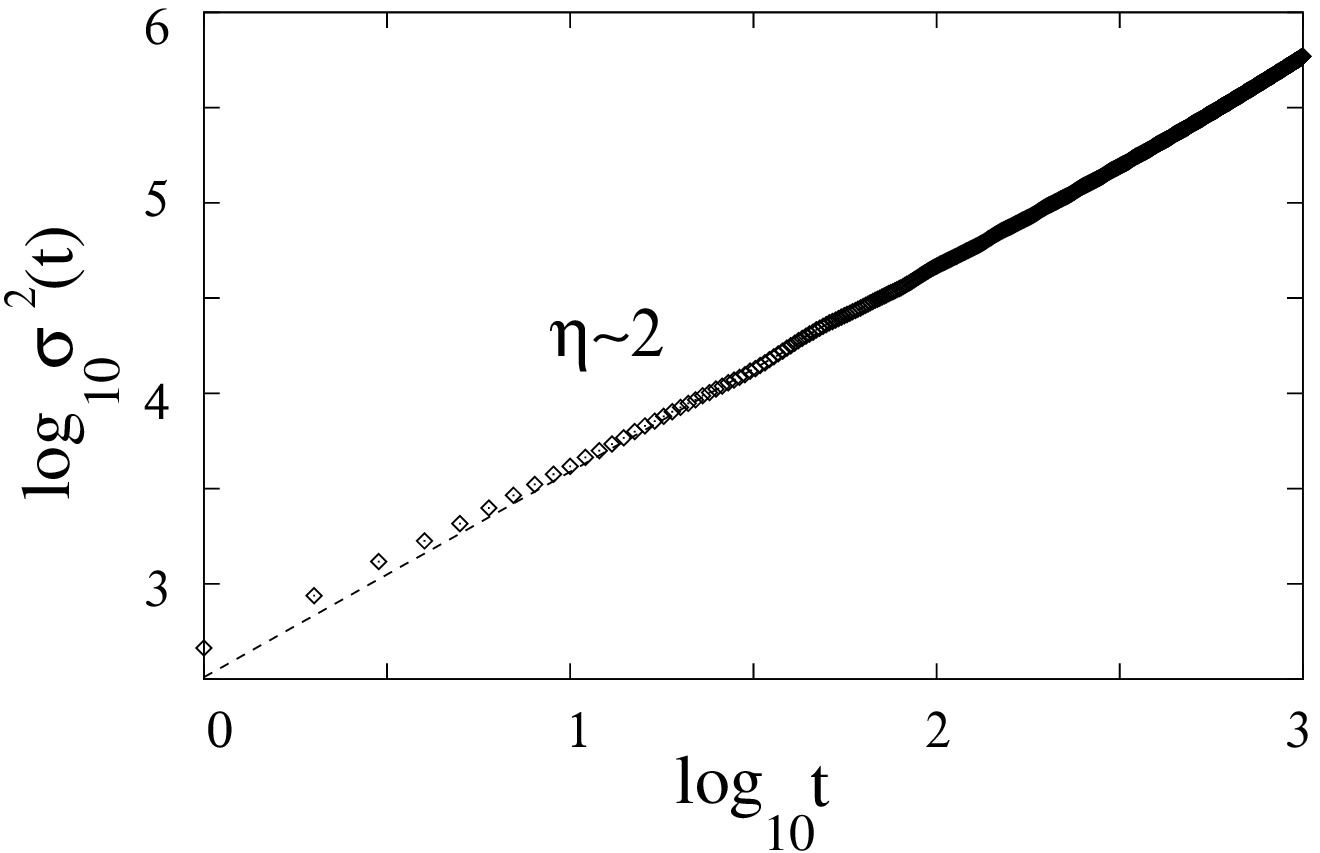}}&
\hspace{0.8cm} (b)&
\resizebox{75mm}{75mm}{\includegraphics{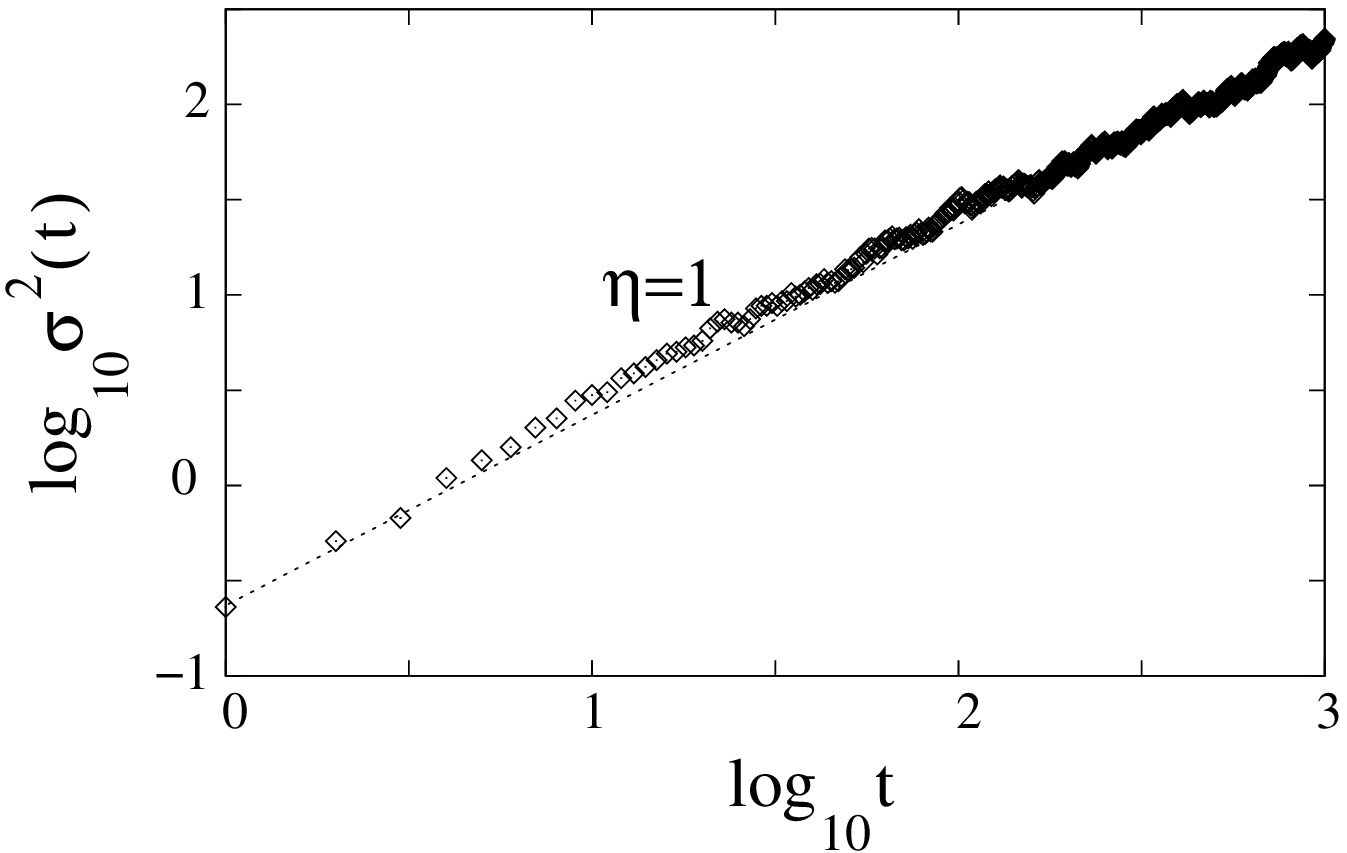}}\\
\end{tabular}
\begin{tabular}{cccc}
\hspace{-0.8cm} (c)&
\resizebox{75mm}{75mm}{\includegraphics{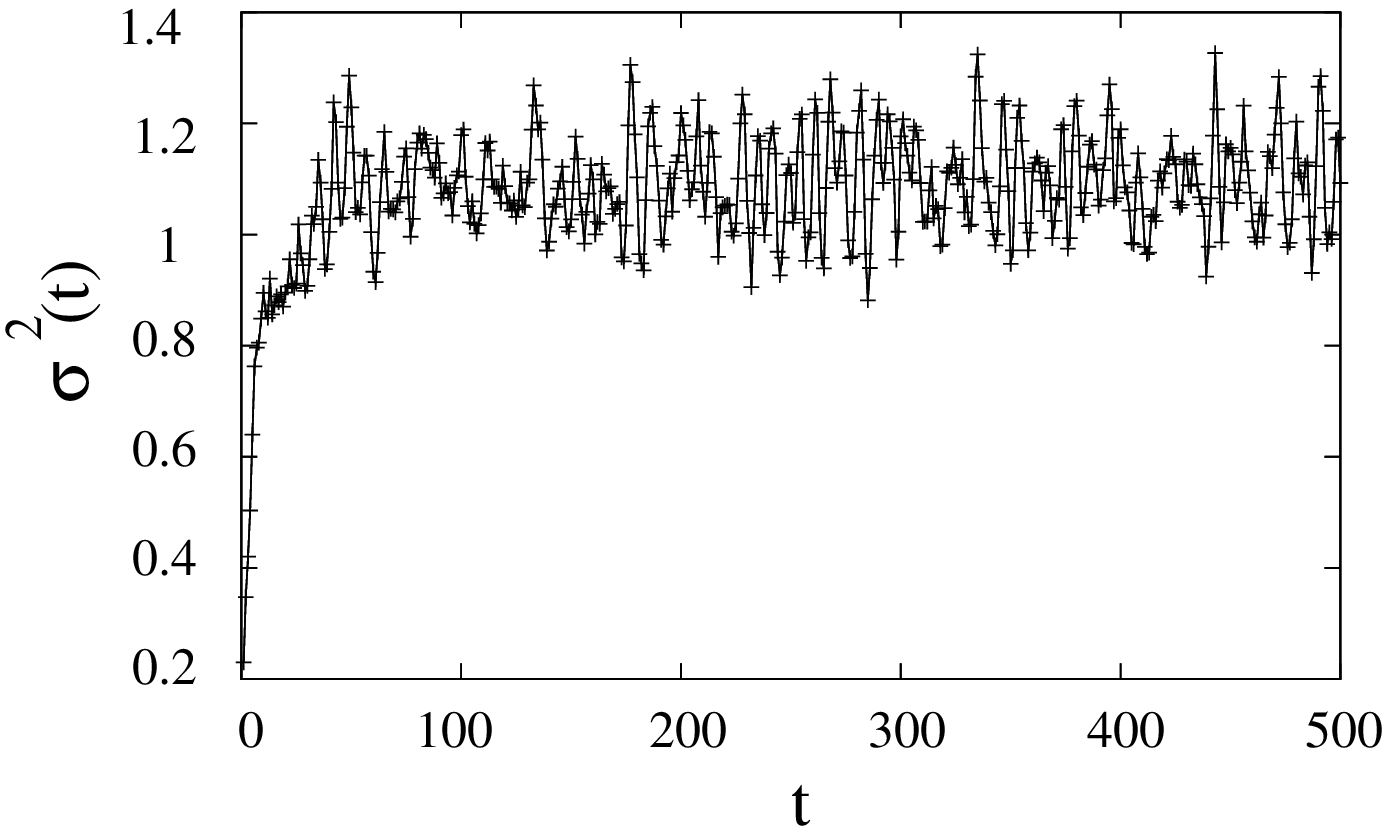}}&
\hspace{0.8cm} (d)&
\resizebox{75mm}{75mm}{\includegraphics{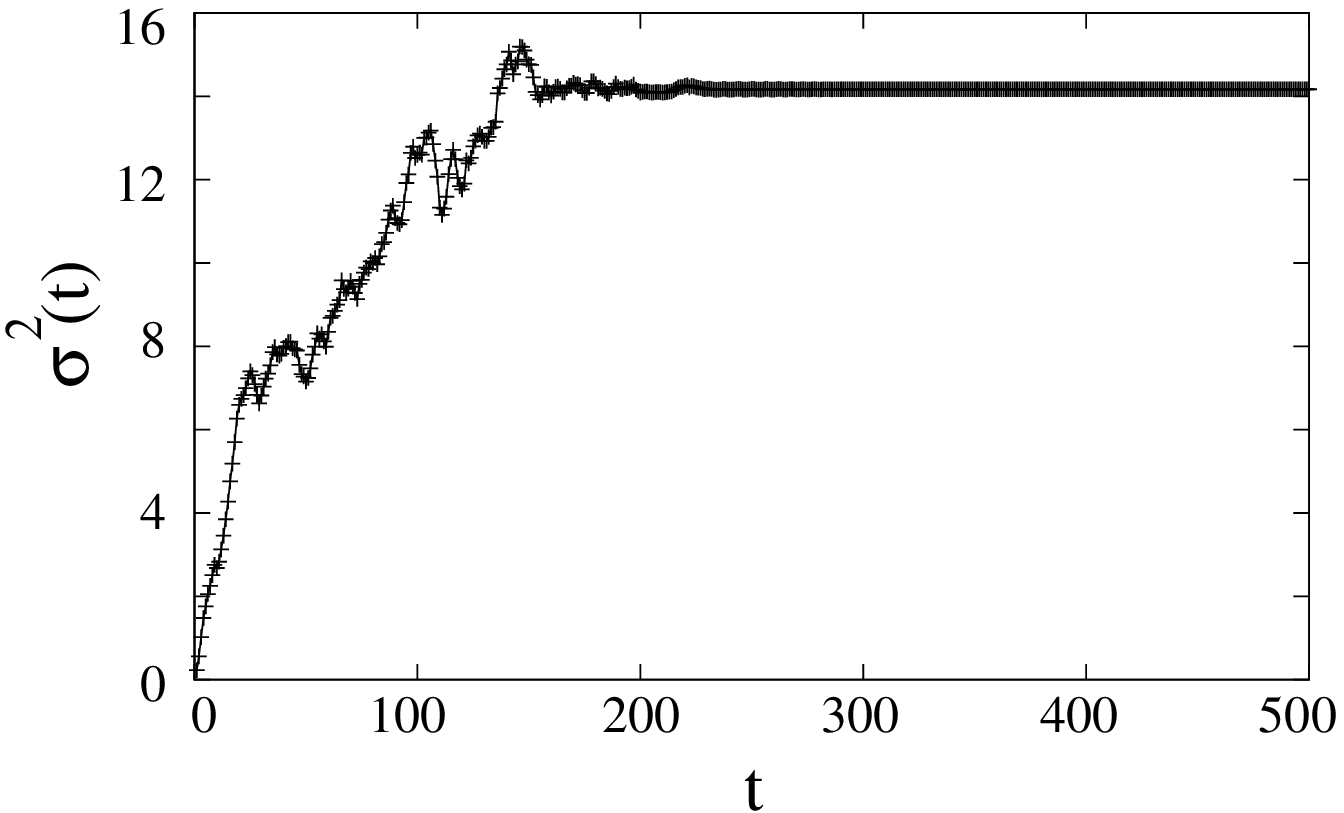}}\\
\end{tabular}
\caption{The plot showing the variance of the particle cloud as it evolves in time.
(a) Superdiffusive behaviour is observed for parameters $\alpha=1.2$ and $\gamma=0.2$ with
exponent  $\eta = 2.0135$.  (b) Normal diffusion is  seen for the parameters $\alpha=2.8$ 
and $\gamma=0.1$ with the exponent $\eta = 1.00611$ (error bars).  Trapping regimes show
the plateauing of the variance  (c) with nonstationary states ($\alpha=0.5$ and $\gamma=0.5$)
and (d)  trapping with stationary states ( $\alpha=1.5$ and $\gamma=0.8$). \label{diff}}

\end{figure}

 Here ${\bf x}(t)$ is the  position  of a particle  and $<{\bf x}(t)>$ is the average position of all the particles at time t, both in the configuration space.  The diffusion  coefficient $D$ and the exponent $\eta$ quantify the type of  diffusion. The angular brackets indicate the average over the ensemble. The configuration space in two dimensions considered here is the cover space, i.e., without the periodic boundary condition that was used in the previous section. Generally, when the exponent in Eqn. \ref{eqn-diff} takes values $\eta>1$, the process is called superdiffusion  and the trajectories of the particles have long displacements. The transport is characterized 
 by normal diffusion if the variance grows linearly with time, i.e. if $\eta =1$. Subdiffusive transport occurs 
 when  the exponent $\eta $ takes values less than 1.  The embedding map shows all the three main classes of diffusion processes as illustrated in Fig. \ref{diff}. Subdiffusive behaviour in the embedding map can be further classified into two subclasses, viz., one associated with the  trapping regions with  non-stationary states and the other with the trapping regions with stationary states. We analyze the behaviour of these classes in detail below.

The phase diagram of Fig. \ref{3dreg}  separates out the distinct dynamical regimes in the $\alpha-\gamma$ space.
Here, we find it  useful  to obtain a phase diagram which classifies  the main diffusion regimes
as a function of the parameters $\alpha-\gamma$. The value of the exponent $\eta$ can be used for
such a classification.   For this,  the log-log plot of the variance as a function of time is  fitted
to a straight line after the initial transients. We used a linear  square fit  for finding the value of
$\eta$ for each data point in the $\alpha-\gamma$ parameter space.

\begin{figure}
\begin{tabular}{cccc}
\hspace{-0.8cm}(a)&
\hspace{-0.8cm}\resizebox{75mm}{75mm}{\includegraphics{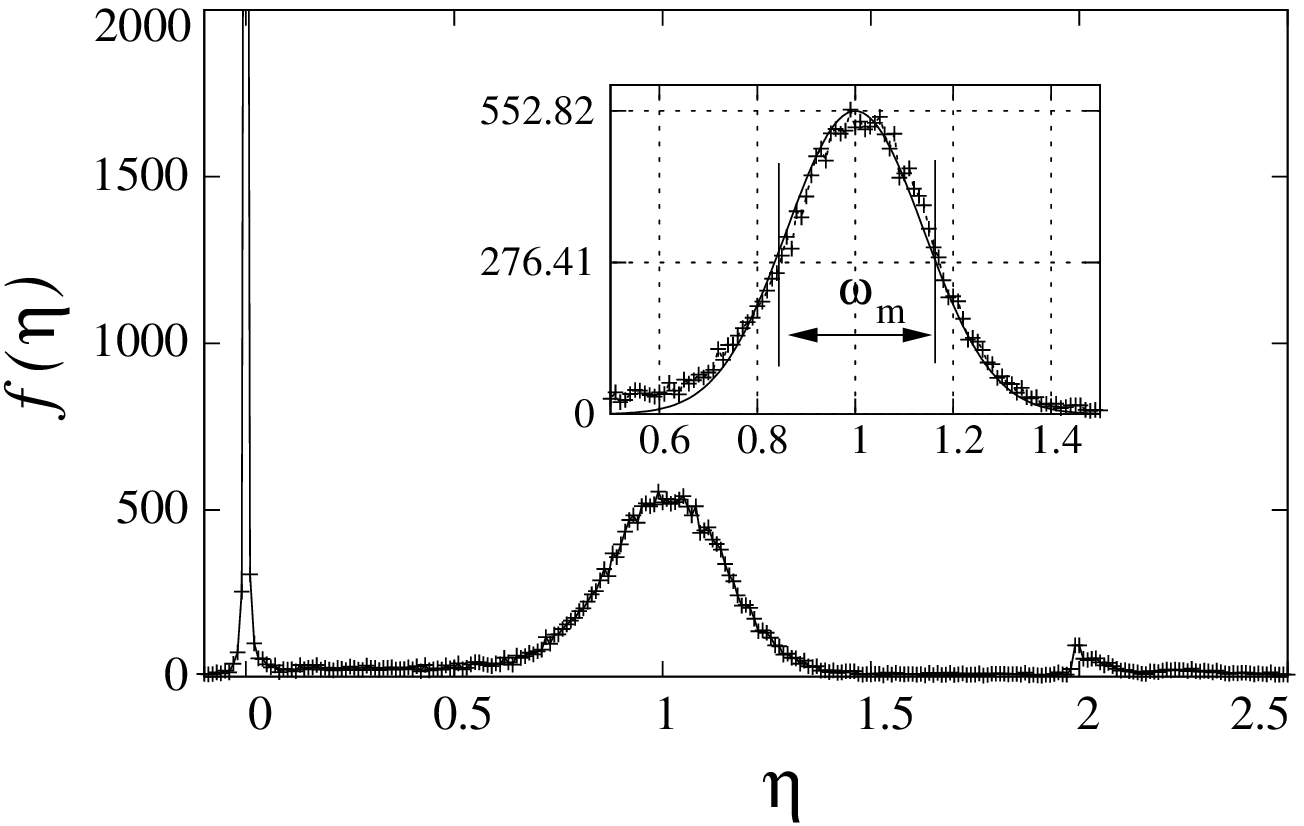}}&
\hspace{0.8cm} (b)&
\resizebox{75mm}{75mm}{\includegraphics{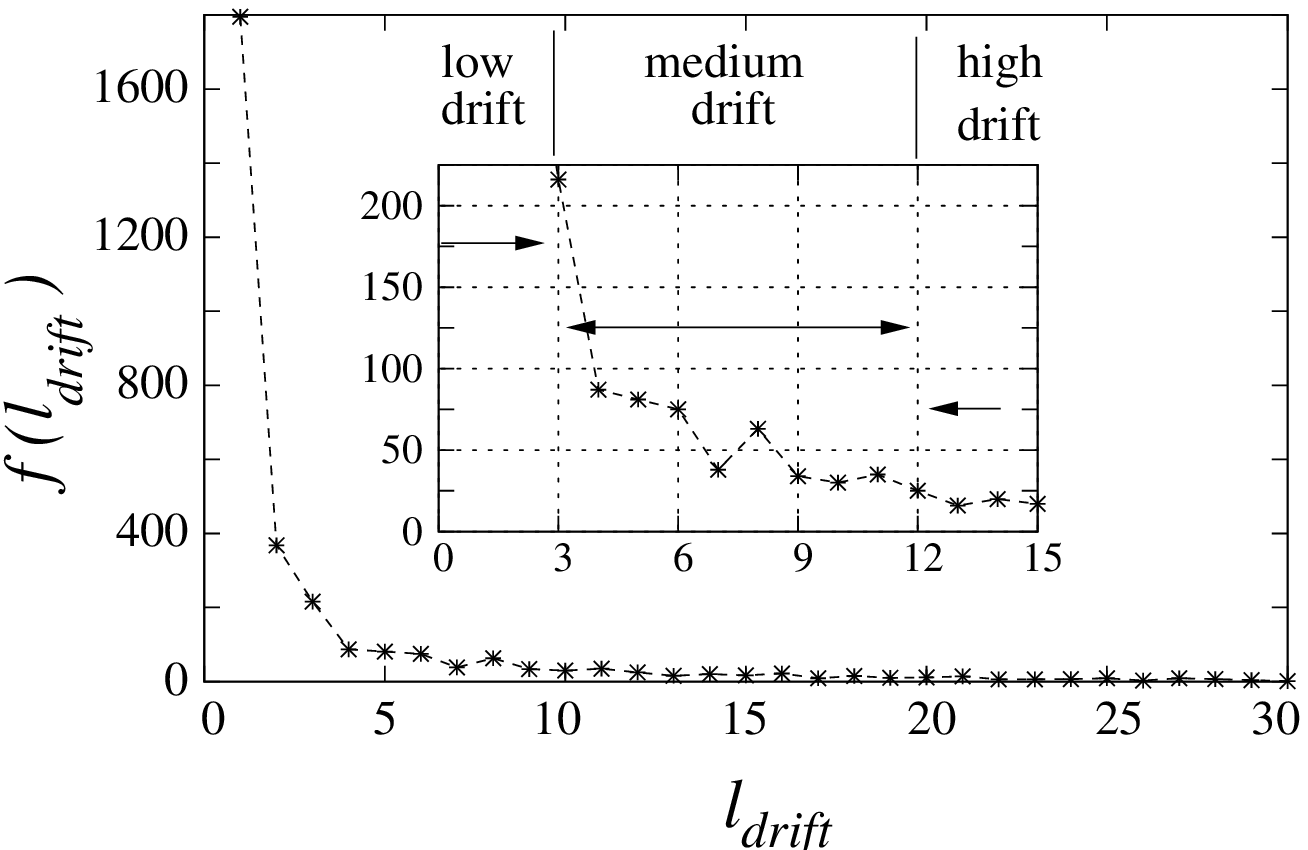}}\\
\end{tabular}

\caption{  The probability distribution of the exponents $\eta$ in the phase diagram, showing three clear peaks. The inset shows the peak centered around $\eta =1$ for which the full width at half maximum  $\omega_{m}$ is in the range  $0.835< \eta < 1.165$ that will delimit the normal diffusion exponents. The peak value at $\eta=0$ is $f(\eta)=6000$. (b) The probability distribution of drifts in the phase diagram has three classes, viz., high drifts ($\overline{l_{drift} } > 12$), medium drifts ($12 > \overline{l_{drift} } > 3$), and low drifts ($3 > \overline{l_{drift} }$) (see inset).  \label{class}}
\end{figure}

Theoretically,  the normal diffusion regions in the phase diagram can be distinguished  from the 
anomalous diffusion regions if the value of the exponent $\eta > 1$ and $\eta < 1$. When dealing 
with numerical data, the normal diffusion regions cannot be identified by selecting the ones with
the exponent $\eta =1 $.  The distribution of the values of $\eta$ in the entire phase diagram is shown in 
Fig. \ref{class}(a), and shows  three peaks, centered around $\eta \sim 0$, $\eta \sim 1 $  and $\eta \sim 2$.
The inset shows the distribution centered around  $\eta \sim 1$ that can be  fitted with  a Gaussian.
The standard deviation of the Gaussian is $\sigma_{gauss} = 0.14$, and the full width at half maximum ($\omega_m$) for 
this Gaussian, expressed as   $\omega_m$ = $2 \sigma_{gauss} \sqrt{2 log(2) } = 2.3548 \sigma_{gauss}$, 
turns out to be  $0.33$ (see the inset of the  
Fig. \ref{class}(a)).  This $\omega_m$ centered 
around $\eta=1$  can therefore be taken as a good practical bound on the $\eta$, to demarcate
normal and anomalous diffusion regimes. The upper and lower bounds on the normal diffusion region are $\eta_{lower}=1- \omega_m/2$ and
$\eta_{upper}=1+\omega_m/2 $ respectively. The exponent  values  that demarcate the regions of 
the normal diffusion are therefore $\eta_{normal} \in [\eta_{lower},\eta_{upper}]$. 
The cases which have values less than the lower bound $\eta_{lower}$
 are identified as subdiffusive and the ones which have values higher than the upper bound $\eta_{upper}$ 
are identified as  superdiffusive.

\begin{figure}

\begin{tabular}{cccc}
\hspace{-0.8cm}(a)&
\hspace{-0.8cm}\resizebox{75mm}{75mm}{\includegraphics{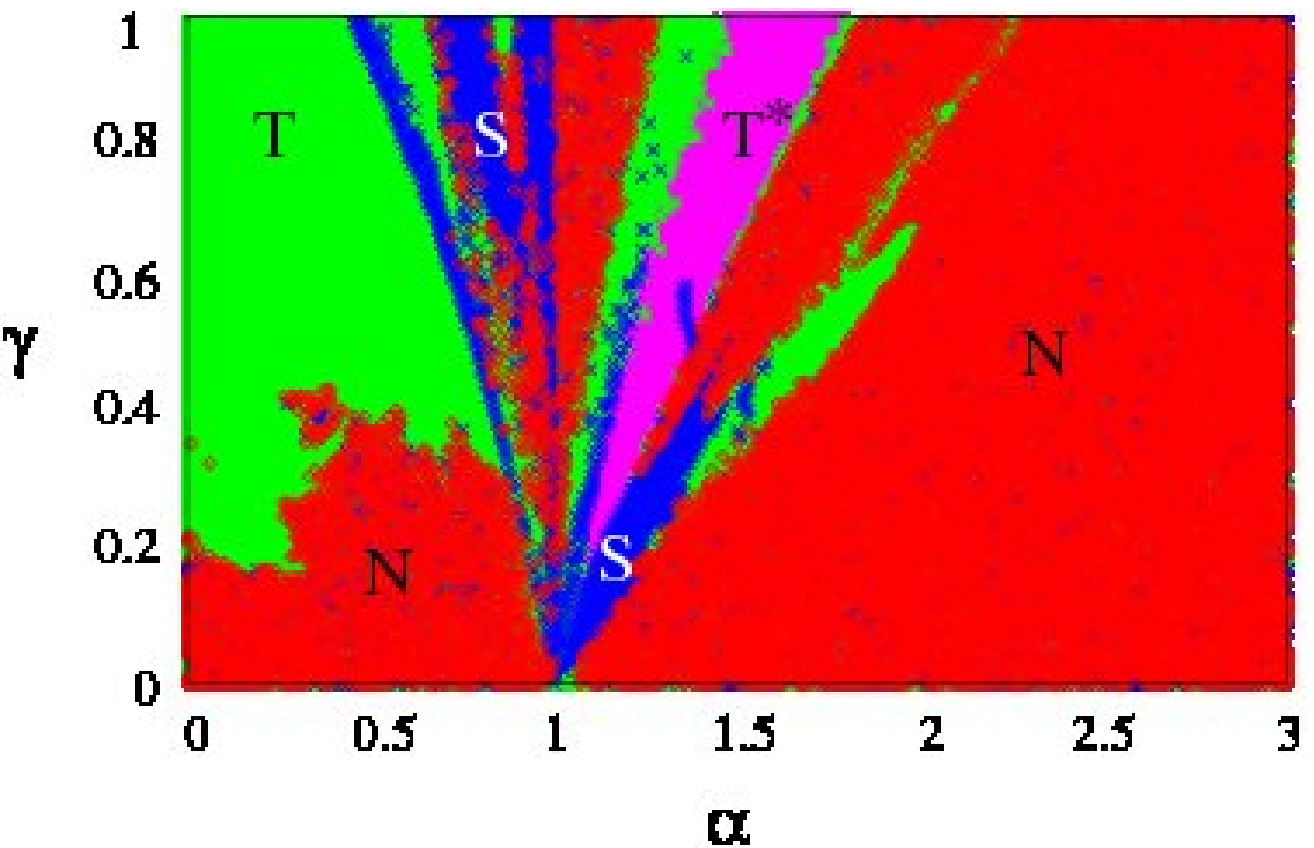}}&
\hspace{0.8cm} (b)&
\resizebox{75mm}{75mm}{\includegraphics{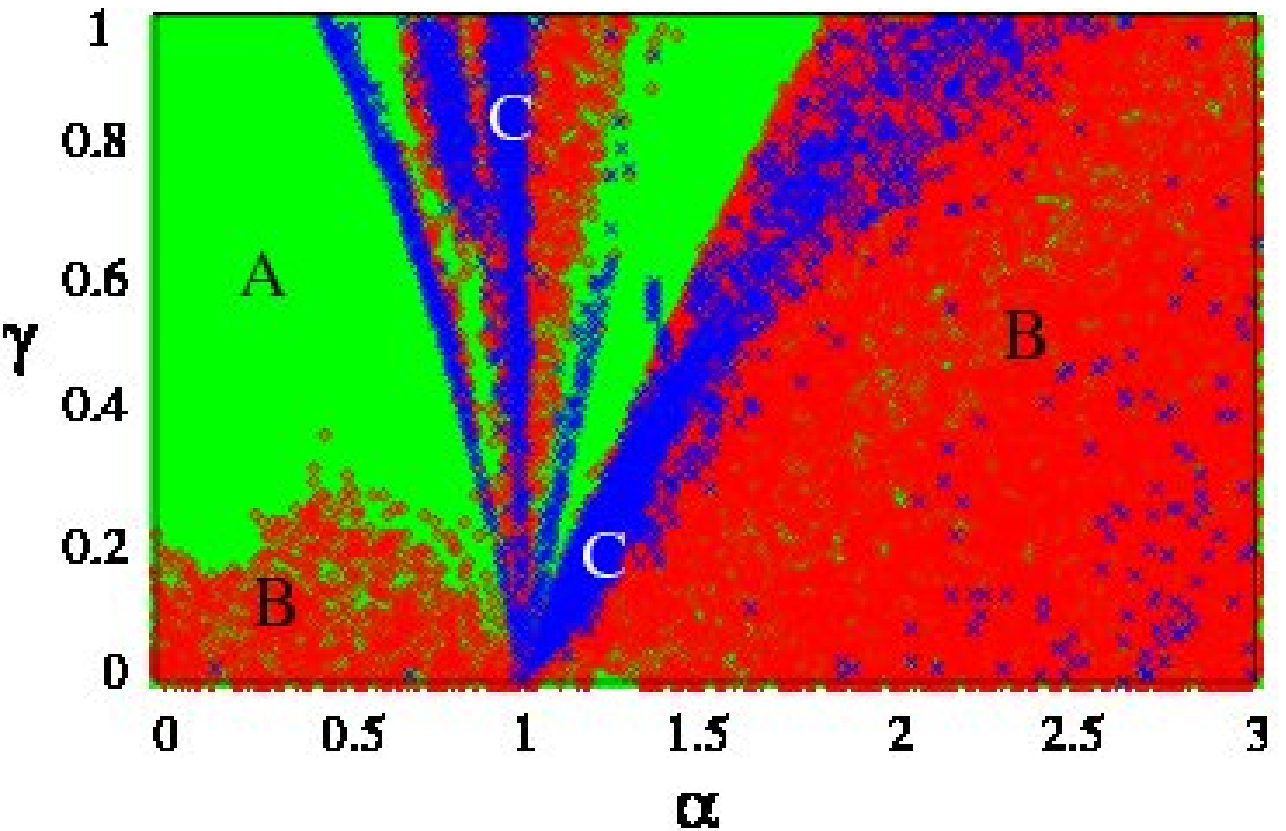}}\\
\end{tabular}

\caption{(Color online) (a) The phase diagrams showing the four diffusion regions in the embedding map in the $\alpha$-$\gamma$ parameter space,  superdiffusive regime is marked by `S' (blue),  normal diffusive regimes are marked by 'N' (red), subdiffusion with stationary states are marked by `T*' (pink) and  subdiffusion with nonstationary states are marked by `T' (green). (b) The phase diagram  shows the drifts of particles in the $\alpha$-$\gamma$ parameter space. The regions marked in `blue'  have large average drifts, $ \overline{l_{drift} } > 12$ (labelled by `C'), regions in `red' have medium average drifts,  $12 >  \overline{l_{drift} } > 3$ (labelled by `B'), and regions in `green' have low drifts,  $ \overline{l_{drift} } < 3 $ (labelled by `A'). This is for $3000$ time steps averaged over 100 random initial conditions uniformly distributed  in phase space.   \label{diff_drift}}
\end{figure}

We note that in generating the phase diagram for diffusion regimes we 
have used  3000 time steps with 100 initial conditions distributed uniformly
 randomly in the phase space. For the  phase diagrams in Fig. \ref{diff_drift}
 a  resolution of  $0.01$  in both the directions was used.

The diffusive phase diagram indicates  the following transport properties. The phase diagram shown 
in Fig. \ref{diff_drift}(a) has normal diffusion regions in the regions that are labelled `N' (colored red), which are  predominantly seen
in the areas where the mixing is seen in the phase space (see Fig. \ref{3dreg}). The superdiffusive
regions are  labelled `S' (coloured blue), as mentioned above, are identified where  $\eta > 1.165$.  
On comparing these regions with  the phase diagram of the dynamical  regimes, it is clear that the  superdiffusive  regions coincide largely  
with  the periodic regime.

\begin{figure}
\begin{tabular}{cccc}
\hspace{-0.8cm} (a)&
\resizebox{75mm}{75mm}{\includegraphics{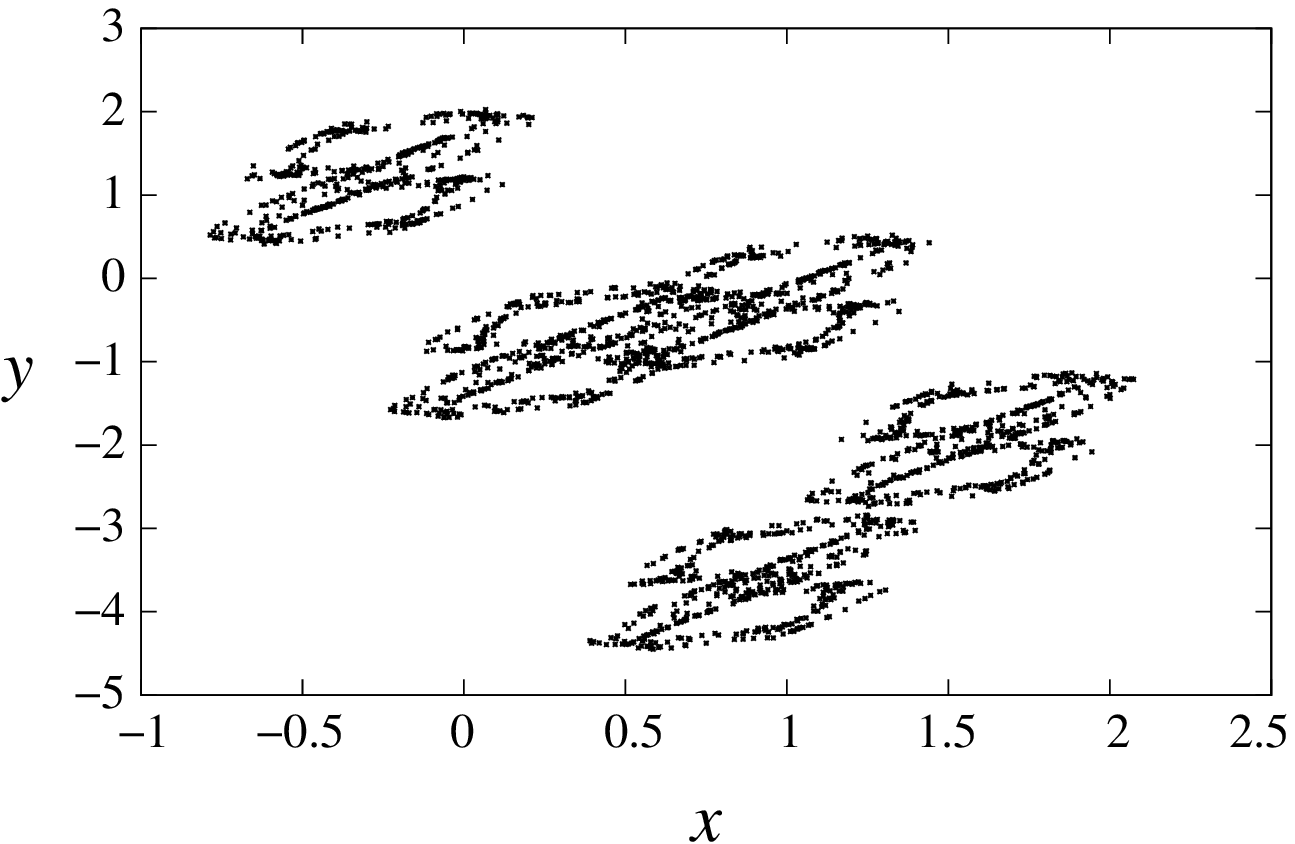}}&
\hspace{0.8cm} (b)&
\resizebox{75mm}{75mm}{\includegraphics{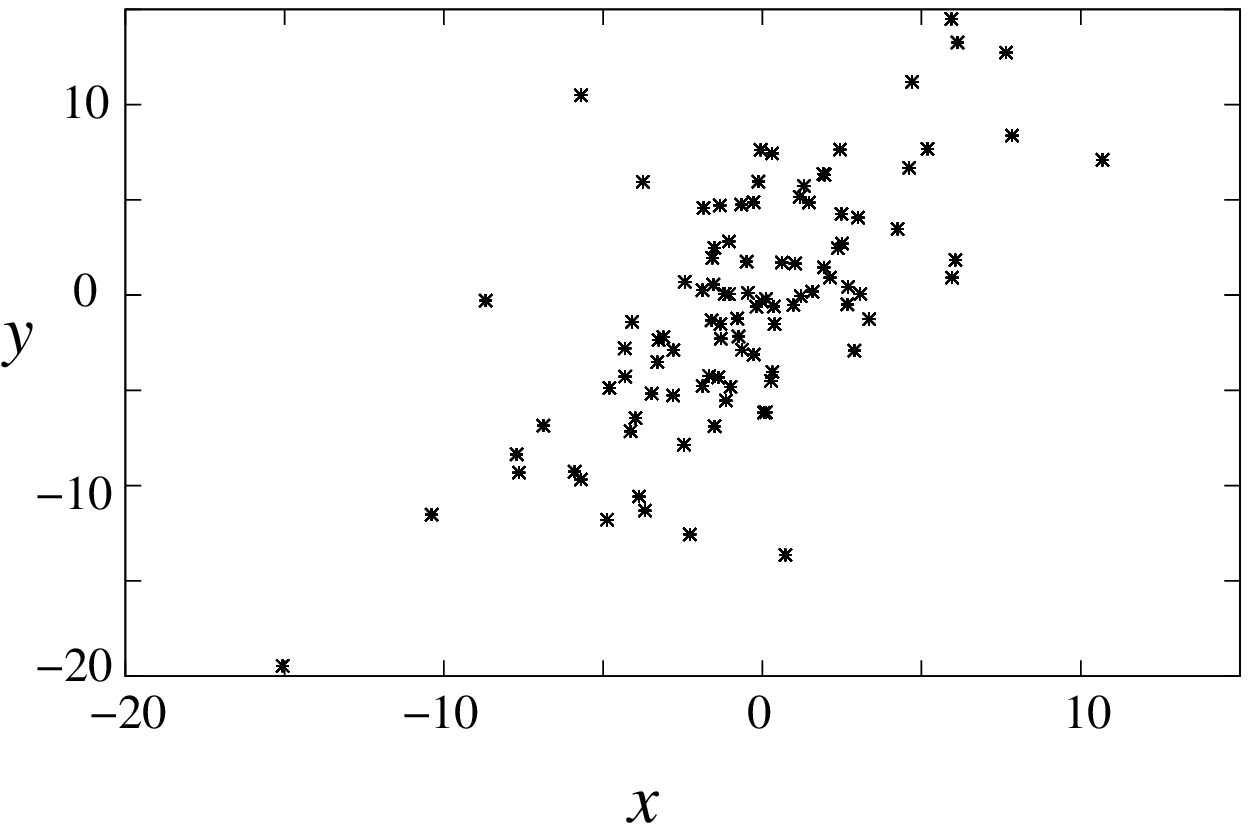}}\\

\end{tabular}

\caption{  The phase space plots of the subdiffusion regimes (a) nonstationary states (b) Trapping  states with stationary points \label{trap}}
\end{figure}

The subdiffusive  region is delimited  by the criterion $\eta < 0.835$. It is easy to see that these regions mostly coincide with the chaotic structures regimes seen in Fig. \ref{3dreg}.  Further, there are two types of subdiffusion occurring here, viz. those associated with trapping in stationary states and  trapping in non-stationary states.  On the aerosol side of the phase diagram   (colored green and labelled `T'), particles behave as though they are trapped in an attractor, but their dispersion grows sublinearly with 
time (Fig. \ref{diff}(c)). The cover phase space for such a typical case is seen in Fig. \ref{trap}(a).
On the other hand, in the bubble region (colored pink and labelled `T*'), particle trajectories get 
trapped forever and become stationary. Fig. \ref{trap}(b) shows the cover phase space for the case 
for trapping into stationary states.  As a result of the motion tending to fixed points, the variance tends
to a constant value without any  fluctuation after the transient time as seen in Fig. \ref{diff}(d).

There are certain parameter regimes at which the particles do not diffuse beyond a given region in the phase space and we saw that they belong to the subdiffusive regions. The average drift of the particles 
 can be used to quantify the overall drift of the particle cloud from the initial position  it started with. The drift is defined as the  mean position of the cloud of the particles after a time t, i.e., $\overline{l_{drift} } = \overline{l(t)} - \overline{l(0)} $, i.e. where 
$\overline{l_{drift} }$ is  the difference between the average position $\overline{l(t)}$ of the cloud of the particles in cover space, 
at time $t$ from that at the time $t=0$. The distribution of the average drifts of the particle cloud for the range of values of $\alpha-\gamma$ discussed in the phase diagram, 
 is plotted in Fig. \ref{class}(b).  It is clear from   Fig. \ref{class}(b) 
that there are three main ranges of drifts. The large drifts are identified  by the criterion  $  \overline{l_{drift} } > 12$,  the intermediate ranges of  drift by $3< \overline{l_{drift} }  < 12$, while the  small drifts are identified to have drift rates $\overline{l_{drift} } < 3$ (see inset of Fig. \ref{class}(b)).

A  phase diagram for the drifts in the $\alpha-\gamma$ space  is plotted in Fig. \ref{diff_drift}(b). Regions of large drift are coloured blue, those of intermediate drift are coloured red, and green regions have small drift. 
On comparing the parameter regions  `T*' (pink) in Fig. \ref{diff_drift}(a)  where particles are trapped and become stationary, with the corresponding region of Fig. \ref{diff_drift}(b),  it is clear that the particles here have low values of drift.  The regions of high drifts in Fig. \ref{diff_drift}(b)
overlap with the superdiffusion regions `S' (blue) both in the aerosol and the bubble regime in Fig. \ref{diff_drift}(a). 
Therefore, there is good correlation between the phase diagrams constructed from the dispersion and drift criteria.  

The phase diagrams obtained in Fig. \ref{diff_drift} show 
no clear contiguous areas defining the three diffusion regimes, or the three drift regimes, and these  regions  are seen to 
be interspersed  in some areas of the phase diagrams. This is due to the fact that the values of the slope $\eta$ and the average drift $l_{drift}$ belong to  continuous distributions which are partitioned into three regions. Hence sharp demarcations are not achieved.

\begin{figure}
\begin{tabular}{cccc}
\hspace{-0.8cm} (a)&
\resizebox{75mm}{75mm}{\includegraphics{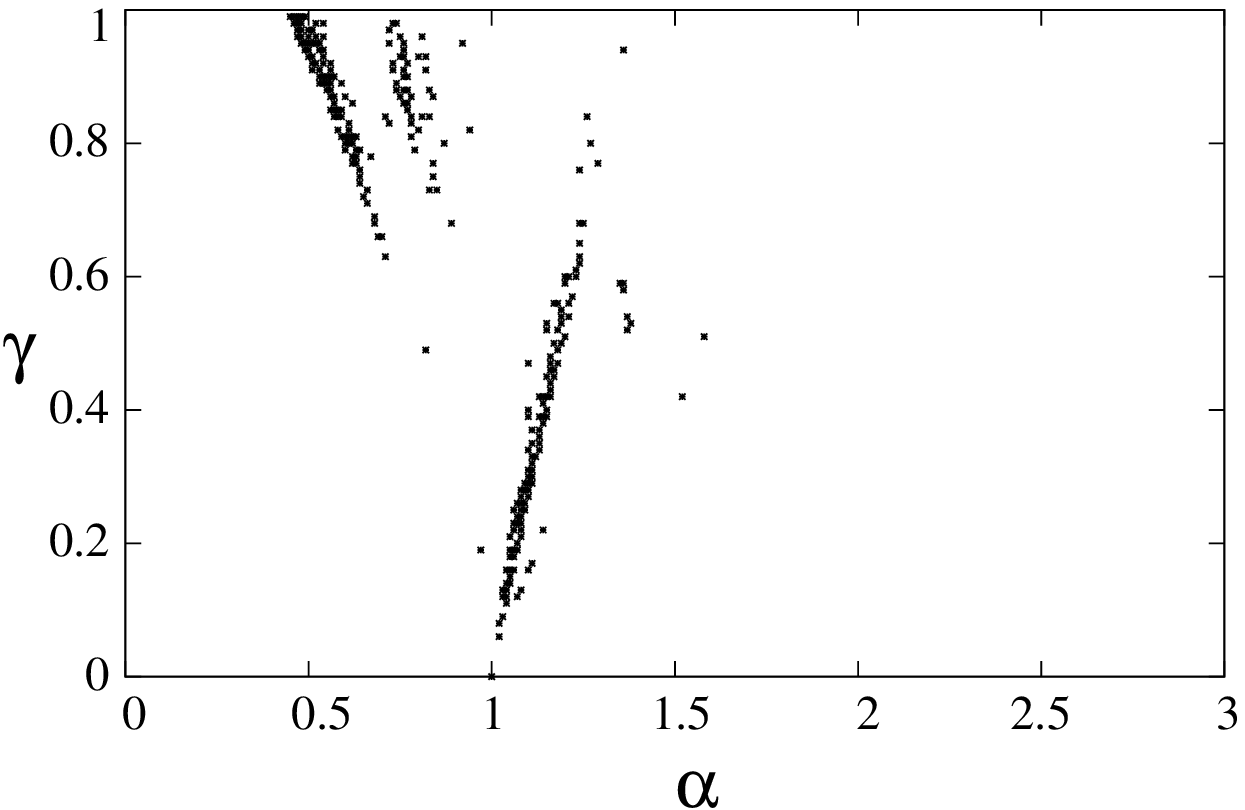}}&
\hspace{0.8cm} (b)&
\resizebox{75mm}{75mm}{\includegraphics{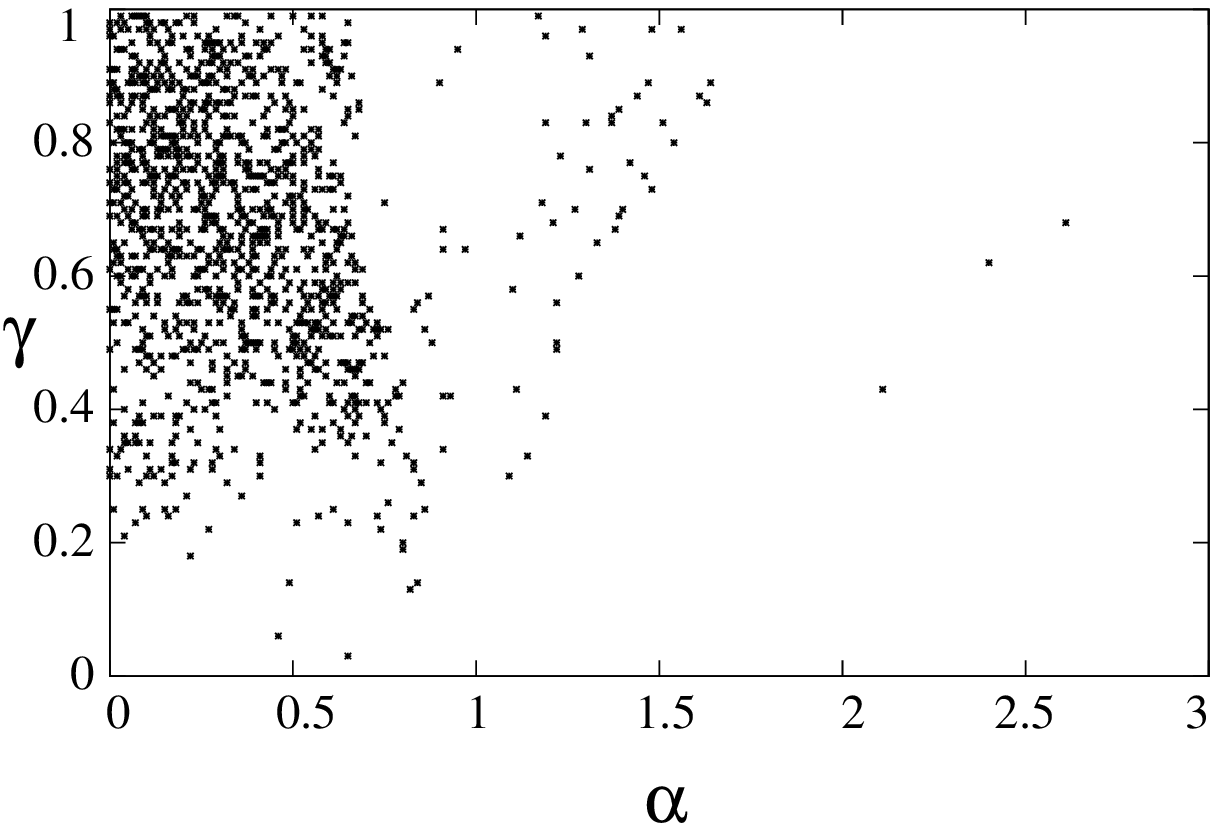}}\\

\end{tabular}
\caption{(a) The phase diagram  shows  the regions characterized by ballistic diffusion. Here the diffusion exponent take values $\eta =2$. (b) The phase diagram with points marked showing regions where the average drift $l_{drift} \sim 0$.   \label{ballistic}}
\end{figure}

We  note that the embedding map also shows a region corresponding to the ballistic superdiffusion regime \cite{zaslav_physrep}
 which is characterized by a diffusion exponent $\eta \sim 2$
and has been earlier seen in diverse contexts such as 
 the motion of atoms, molecules and clusters on solid surfaces \cite{lacasta04} and 
in random walk models with random velocities \cite{zaburdaev08}.  
This regime is shown in the 
phase diagram Fig. \ref{ballistic}(a). We note that this regime is seen at 
the edges of the periodic tongues.  It is also useful to have a phase diagram to 
locate the  regions where the average drift is zero, $l_{drift} \sim 0$. 
The Fig. \ref{ballistic}(b) shows the regions of near zero drift and in reference
to Fig. \ref{diff_drift}(b), we can see that it forms a subset of the regimes of 
the slowest drifts, marked in green (labelled `A').

It is important to note that characterization and classification of different diffusion and drift
regimes are very useful  in real application contexts. We expand on this point in the conclusion.

\section{The behaviour of the initial transient}

The phase diagrams of the previous section have been obtained after  the systems has equilibrated after an initial transient. 
However, the behaviour of the initial transient can itself be quite different in the case of  distinct dynamical regimes. We illustrate this for the dynamical regimes of the embedding map.

\begin{figure}
\begin{tabular}{cccc}
\hspace{-0.8cm} (a)&
\resizebox{75mm}{75mm}{\includegraphics{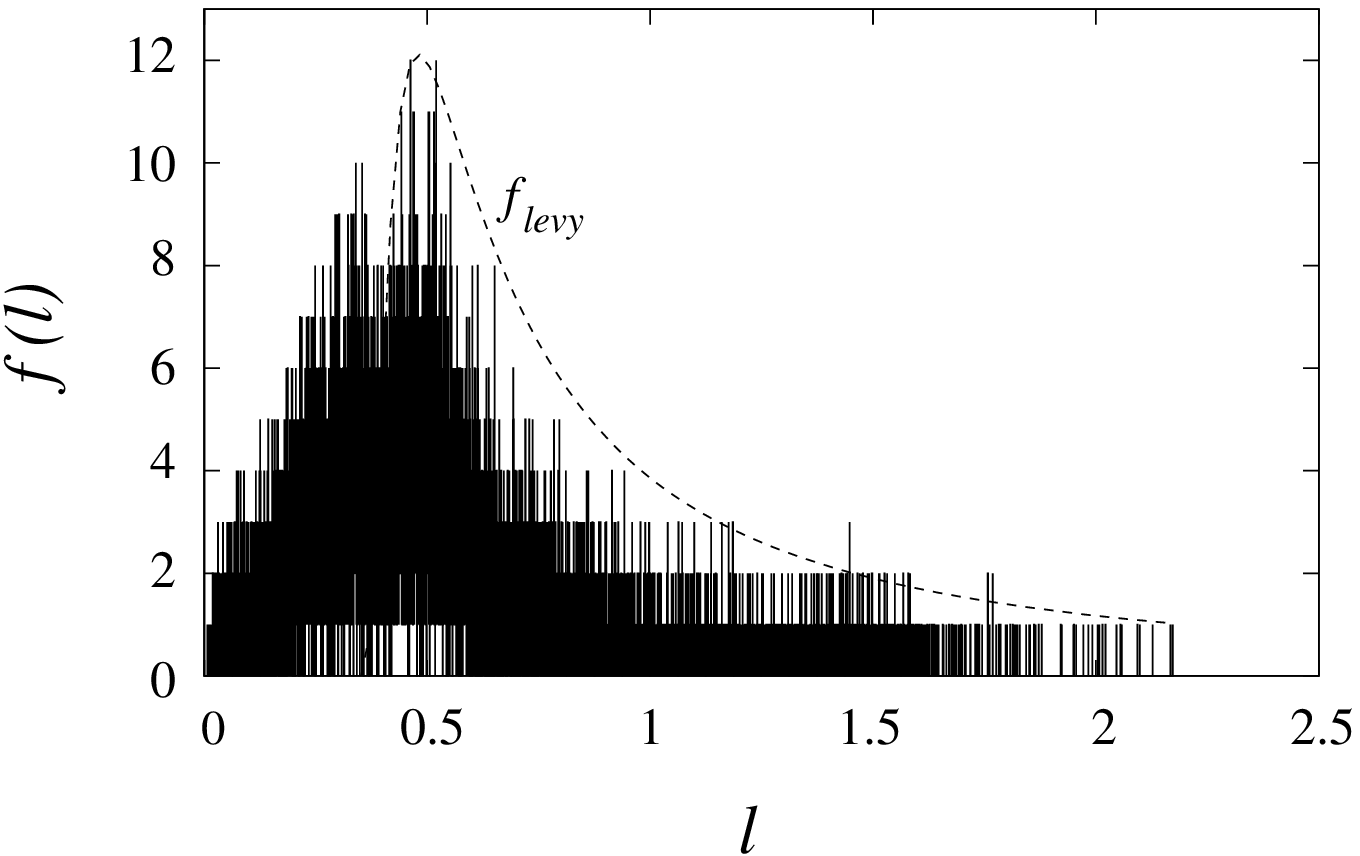}}&
\hspace{0.8cm} (b)&
\resizebox{75mm}{75mm}{\includegraphics{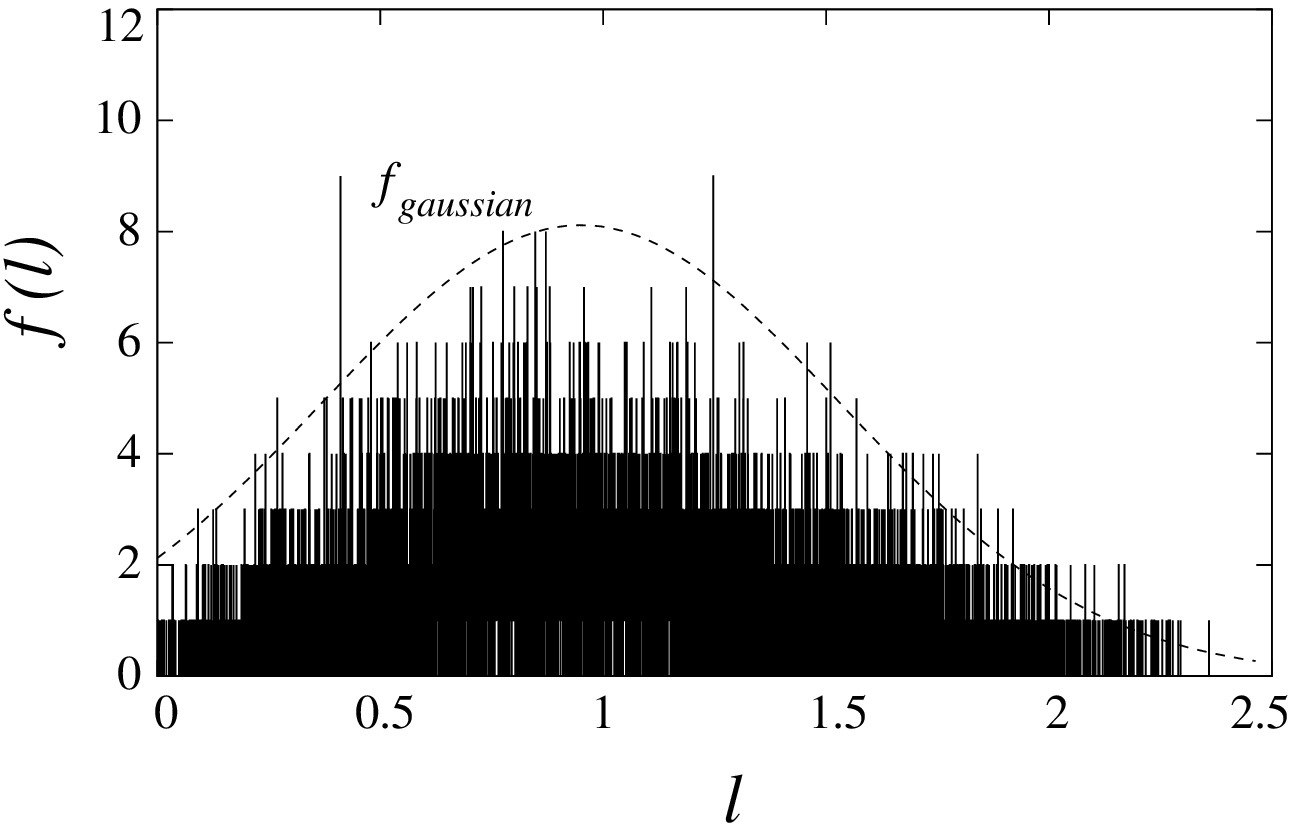}}\\

\end{tabular}

\caption{  The envelope of the distribution of flight  lengths of trajectories in the periodic and chaotic structure regimes is fitted with a heavy tailed distribution. (a) the periodic regime ($\alpha=0.5$,$\gamma=0.5$). (b) In the mixing regime the envelope is fitted by a Gaussian ($\alpha=2.8$,$\gamma=0.1$). In both the plots 8000 bins were used for averaging in the x-axis. )\label{levy}}
\end{figure}

We examine the  probability distribution of the jump lengths, i.e. the Euclidean distance between successive iterates,  for transient times  upto $500$ iterates in regimes corresponding to  periodic, chaotic and mixing behaviour. Here, $1000$  initial conditions are spread uniformly randomly in the phase space for this analysis.  Fig. \ref{levy}(a) shows the probability distributions of the jump lengths of the trajectories in the regime ($\alpha=0.5$,$\gamma=0.5$), where chaotic structures are seen in the phase space. The distribution is clearly characterized by a heavy tail. Similar heavy tails characterize the jump length distribution of the trajectories seen at other points in the chaotic structure regime. In the region of the heavy tail the envelope can be fitted by the function,

\begin{eqnarray}
f_{levy}(l) = \kappa  \sqrt{\frac{\nu}{2 \pi}} \frac{1}{(l-\delta_l)^{3/2}} exp \left( \frac{-\nu}{2(l-\delta_l)} \right),
\end{eqnarray} 

where, the parameters take the values $\nu=  0.42$ and $\delta_l=0.34$  and the scale factor is $\kappa=11$ for the case shown in Fig. \ref{levy}(a). On the other hand, the envelope of the distribution of jump lengths of trajectories for parameter values in the mixing regime $\alpha=2.8$, $\gamma=0.1$ shows  Gaussian behaviour.  In the Fig. \ref{levy}(b) the envelope of the distribution has been  fitted with a Gaussian function, $f_{gauss}(l) = \frac{1}{\sigma \sqrt{2 \pi}} exp(-\frac{{(x-\mu)}^2}{2 {\sigma}^2})$, with $\mu=0.95$, $\sigma=0.58$.

Thus, in the initial transient in the chaotic structure regime, the distribution of jump lengths peaks at short values, but decays with a power-law tail, characteristic of the sticky regions. 
On the other hand, the jump length distribution in the mixing regime, shows normal behaviour, indicating rapid mixing of the initial conditions.

\begin{figure}

\begin{tabular}{cccc}
\hspace{-0.8cm} (a)&
\resizebox{75mm}{75mm}{\includegraphics{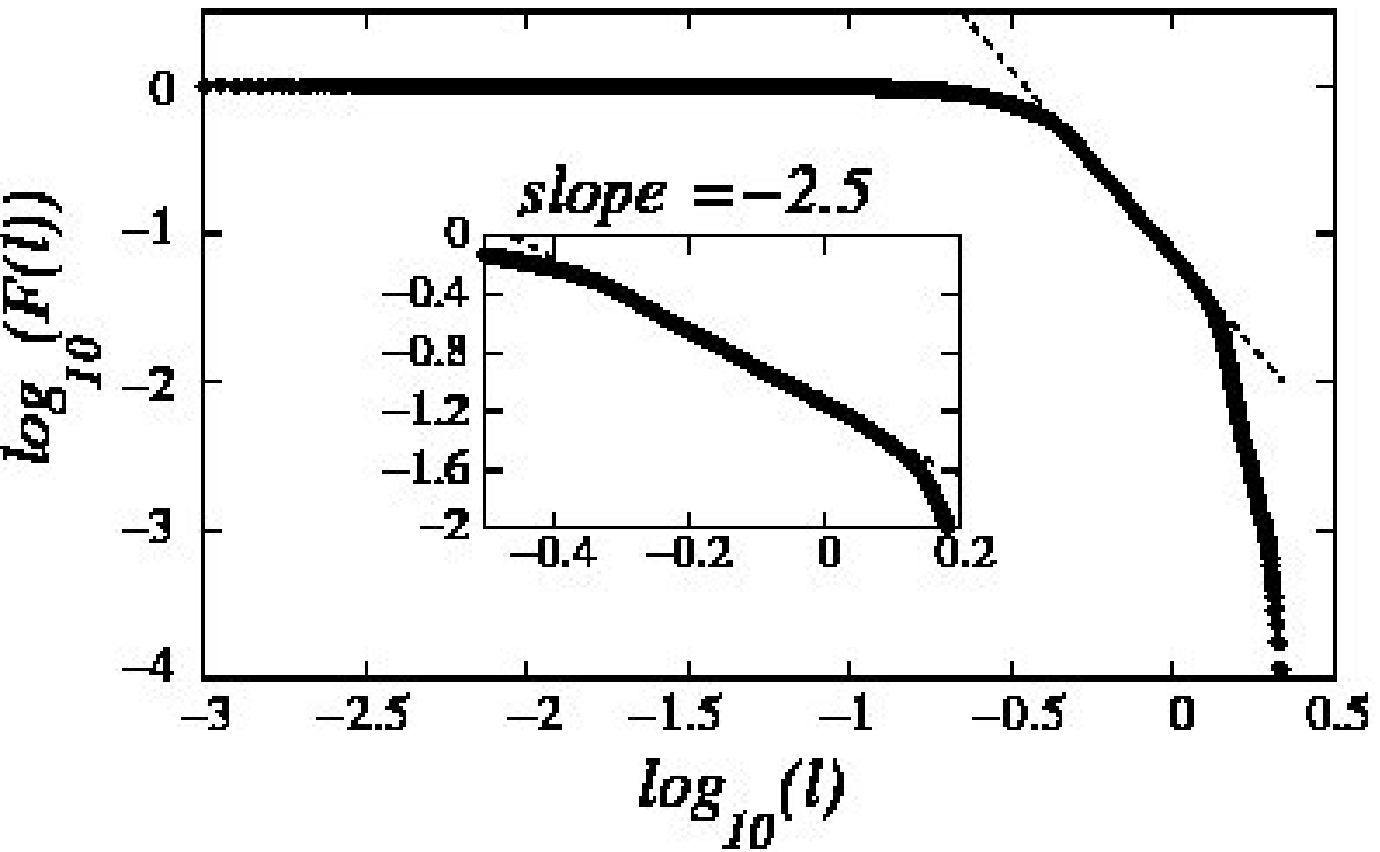}}&
\hspace{0.8cm} (b)&
\resizebox{75mm}{75mm}{\includegraphics{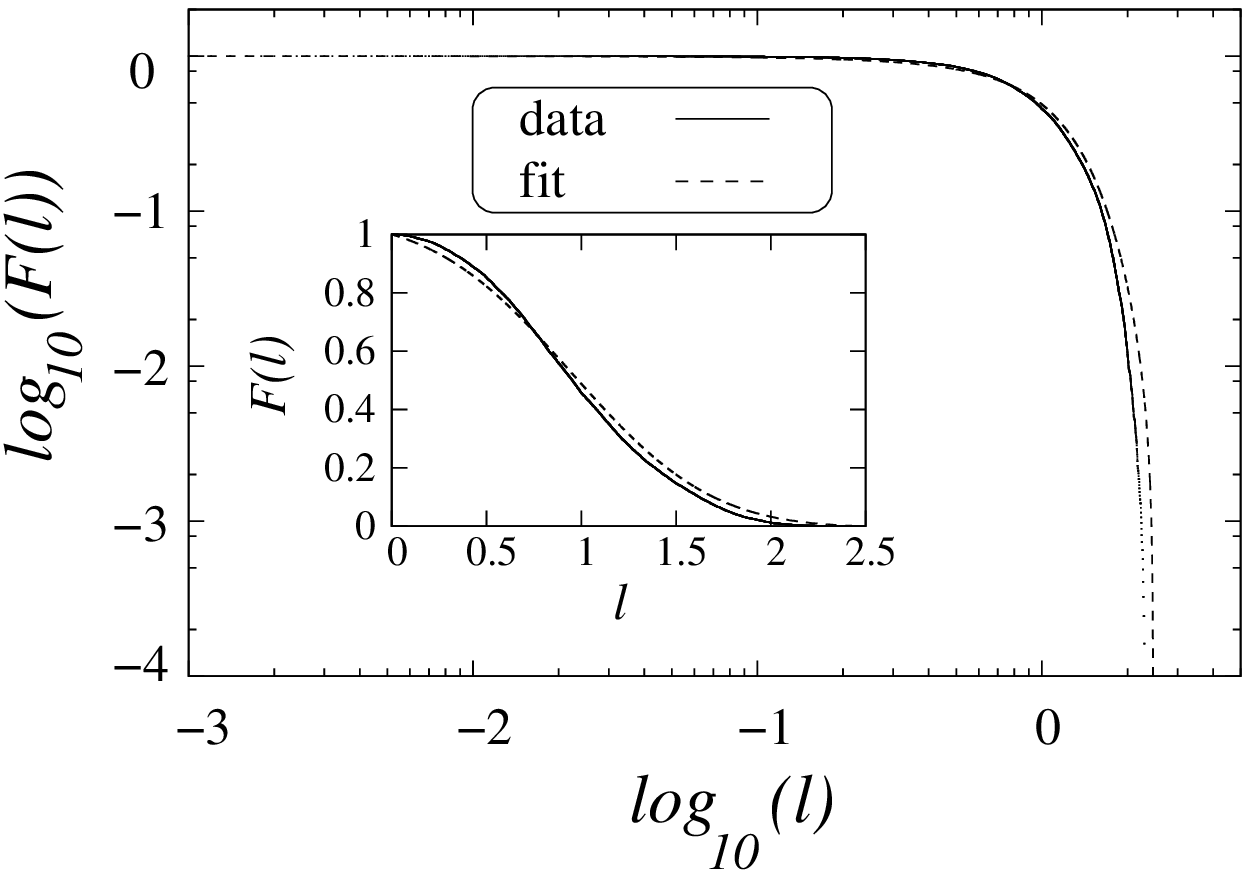}}\\
\end{tabular}
\caption{ The log-log plots of the  cumulative probability distributions of the normalized jump lengths in  (a) a typical  trapping region ($\alpha = 0.5 $, $\gamma = 0.5$) showing a power law tail with slope  - 2.5 and (b) a typical normal diffusion region  ($\alpha = 2.8 $, $\gamma = 0.1$) showing the fit with  a  cumulative gaussian distribution. In both the figures, 8000 bins were used for averaging. \label{ccdf_levy}}
\end{figure}

The cumulated jump distributions show this very clearly. The cumulated 
jump length distributions in the chaotic structure regime show power-law behaviour (see Fig. \ref{ccdf_levy}(a)), whereas the distributions in the mixing regime conform to a  cumulated Gaussian (see Fig. \ref{ccdf_levy}(b)).  The same function $f_{gauss}(l)$ given above is used to obtain the  cumulated distribution $F(l)$.

\section{Transport in the UDV region  }

It was reported recently that the embedding map  system studied here experiences a severe form of nonhyperbolicity  in the aerosol regime in the neighbourhood of the parameter values $\alpha=0.8$, $\gamma=0.40$ \cite{nir08}.
A crisis is seen in the system at these parameter values.
The nonhyperbolicity in higher dimensional systems, in general, manifests itself as unstable dimension variability (UDV) with an accompanying breakdown of the shadowing theorem. The signatures of the presence of unstable dimension variability in the embedding map were seen in the fluctuation of the Lyapunov exponent around zero. The presence of UDV was confirmed from the  distribution  of the finite time Lyapunov exponents which showed that the spectrum  is equally distributed in both positive and negative exponents \cite{nir08}. 

\begin{figure}[!b]
\resizebox{75mm}{75mm}{\includegraphics{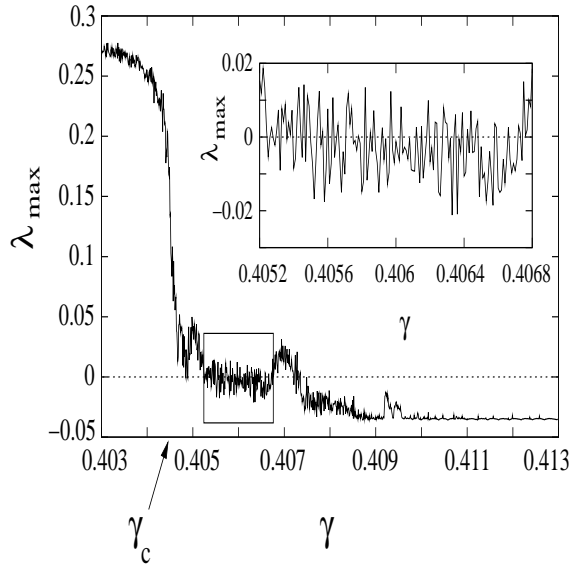}}\\
\caption{The plot of $\lambda_{max}$ vs $\gamma$ showing sudden jump at crisis. The inset shows the fluctuation of 
$\lambda_{max}$ around zero which is a signature of presence of  UDV (fixed  $\alpha=0.8$).\label{lyap_udv}}
\end{figure}
The UDV region is found at an interface between two dynamical regimes, for example in the above case, it is in seen at the interface between the periodic and the chaotic structure regimes. The transport properties of the system are studied in this region.  However, due to the breakdown of 
shadowing at long times, computations  are carried out  for short times. 
Since the region where the UDV is seen lies between the periodic and chaotic structure regime, the statistical properties  such as the recurrence times statistics and the diffusion properties will have contributions  from both 
the periodic and the chaotic structure regime. 

\begin{figure}[!b]

\begin{tabular}{cccc}
\hspace{-10mm}
(a)&
\hspace{-5mm}
\resizebox{75mm}{75mm}{\includegraphics{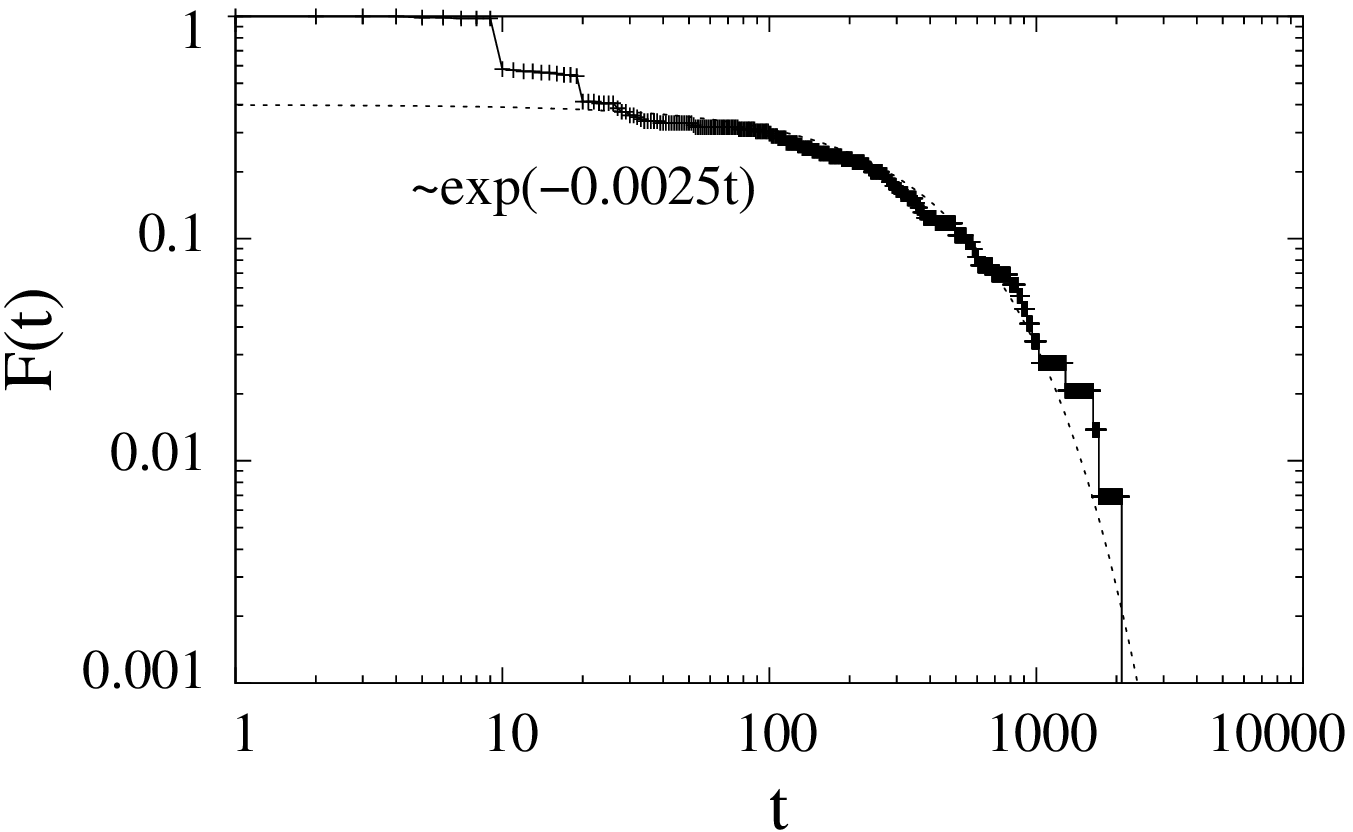}}&

(b)&
\hspace{-5mm}
\resizebox{75mm}{75mm}{\includegraphics{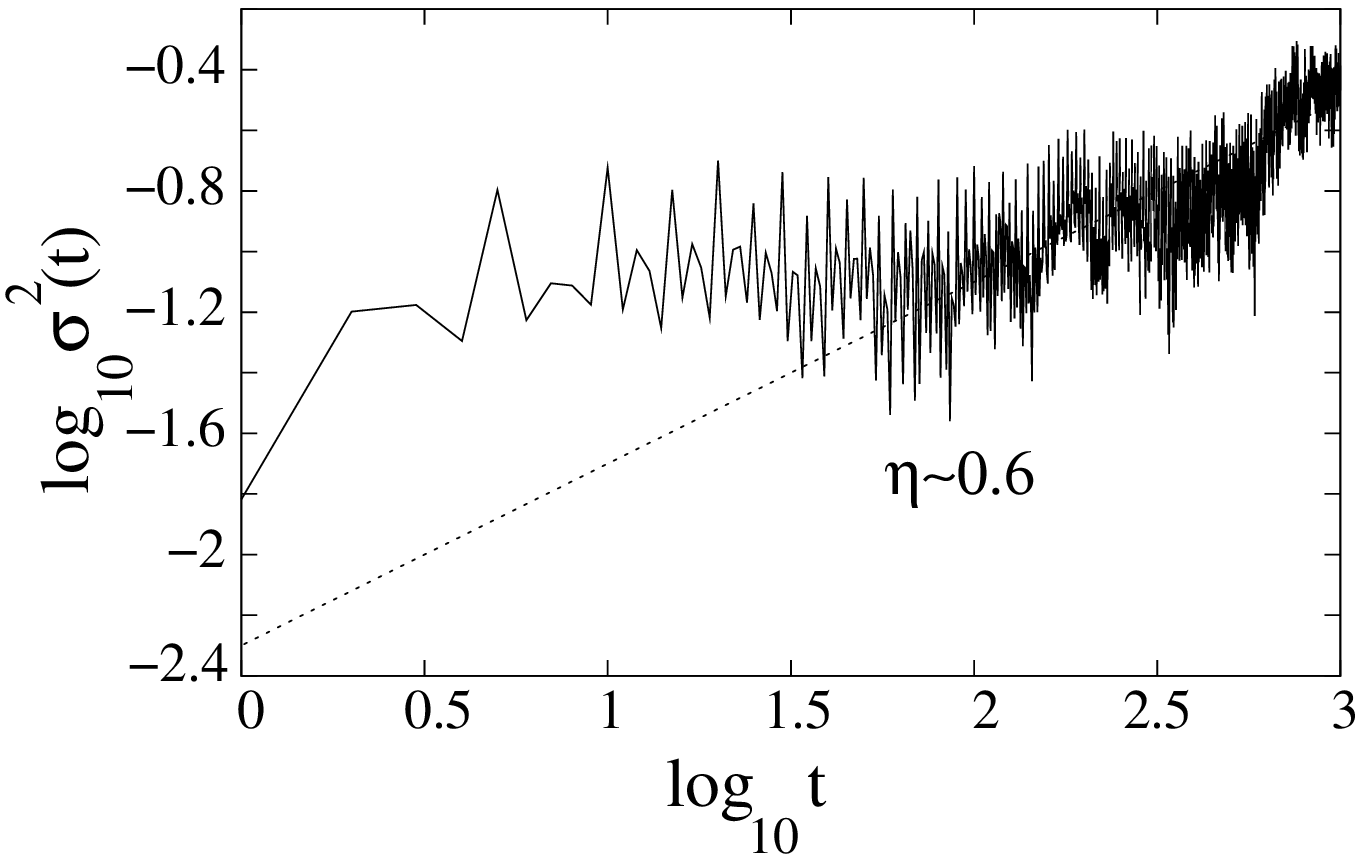}}\\
\end{tabular}
\caption{(a) The cumulative recurrence time distribution showing an exponential fall in the intermediate time range. The discrete plateaus like behaviour in the small and large time scales are due to the periodic nature. (b) The variance as a function of time in the UDV case shows subdiffusvie behaviour. The parameter values are $\alpha=0.8,\gamma=0.4062$.\label{udv2_recdif}}
\end{figure}

The cumulative recurrence time statistics for a typical point in the UDV regime is shown in Fig. \ref{udv2_recdif}(a). There is an exponential decay of recurrence times at intermediate time scales, as seen for the recurrence time distribution in the chaotic structure regime. The periodic regime has its signature at the
short and long  time scales  with the periodic behaviour showing up as discrete steps in the plot. 
The plot of the variance of the particle cloud as a function of time for the same parameters  is seen in Fig. \ref{udv2_recdif}(b). This is clearly a subdiffusive case with the slope $\eta \sim 0.6$, indicating the influence of the chaotic structures.

\section{Connection between the dynamical and transport properties}

It is clear from the previous discussion that there is an intimate connection between the dynamical and statistical properties of the system.
We summarize our inferences in this section.

The phase diagram of the dynamical regimes of the embedding map, a paradigm for the dynamics of inertial particles in fluid flows, contains three distinct 
dynamical regimes, viz. periodic orbits, chaotic structure regimes and mixing behaviour. There is also a clear distinction between the dynamical behaviour in the aerosol regime, and the dynamical behaviour in the bubble regime. Similar 
distinctions are seen in the recurrence time distributions, and the diffusive and drift phase diagrams of the system. 

The recurrence time statistics are useful in the regimes where  no regular periodic behaviour is seen, i.e., in the regions where  the largest Lyapunov exponent is greater than zero. For the embedding map, these are   the chaotic structure regime and the mixing regime. 
The cumulative recurrence time distribution for the chaotic structure regime shows an exponential decay followed by a power law tail. The exponential behaviour at short times reflects the mixing background in the inhomogeneous phase space. The power law seen here is the hallmark of the sticky regions in the phase space. The diffusion studies reveal that the chaotic structure regime shows subdiffusive character on the average. The distribution of  jump lengths in the transient regime has an envelope that conforms to a Levy distribution, indicating that while short jumps dominate the distribution, the number of long jumps is sufficiently large to contribute a power-law tail.

The mixing regime, as expected, shows the exponential decay of the recurrence time distribution. The transport in this  regime shows normal or  brownian  diffusion, where the variance grows linearly with time. Normal diffusive behaviour is predominant in the mixing regime on both the aerosol and bubble sides. The distribution of jump lengths in this regime has a Gaussian envelope.

Two phase diagrams, viz., the diffusion  and the drift phase diagram, were constructed in the $\alpha-\gamma$ parameter space to compare the nature of diffusion and dynamics in all the dynamical regimes of the embedding map. It was found that three main types of diffusion, viz normal (brownian type), sub-diffusive and super-diffusive types are seen in the system. As mentioned above, normal diffusion is the dominant behaviour in the mixing regime on both sides of the phase diagram. Super-diffusive regions are seen in the periodic regime of the aerosols, and in a part of the periodic regime of the bubbles. However, the periodic regime on the bubble side also contains a regime where the inertial particles are trapped and have zero drift. This regime has sub-diffusive transport with stationary states. Here the variance saturates to a constant value after an initial sublinear rise. This regime shows behaviour consistent with early studies of impurity behaviour \cite{eaton94}, where it was observed that bubbles were pushed towards the islands
forming regions of preferential concentration. 

Sub-diffusive transport is also seen in the chaotic structure regime on the aerosol side of the phase diagram where trapped states are also seen. However, the trapped states are non-stationary over here. Here, the variance fluctuates 
about a constant value after the initial sublinear rise.   On the other hand, normal diffusion is predominant in the chaotic structure regime on the bubble side of the phase diagram. Ballistic diffusion similar to that seen in many natural processes is also seen in the embedding map. The regions of ballistic diffusion in the embedding map lie on the boundaries of the periodic regimes. The embedding map also shows the existence of unstable dimension variability  at the boundary between the periodic regime and the chaotic structure regime. Thus, the signature of both the regimes are seen in the recurrence time distribution. However, the transport behaviour is clearly sub-diffusive.

\section{Conclusions}

The transport properties of bailout embedding map, a paradigm to model the dynamics of inertial particles, are studied  in this paper. The base fluid flow is  assumed to be incompressible and modelled by an area preserving map, the standard map. The resulting embedding map is dissipative and captures the qualitative dynamics of both particles that are heavier than the fluid, the aerosols, and lighter than the fluid, the bubbles. The main dynamical regimes in the embedding map system are reviewed and  
the statistical characterizers of transport in each distinct dynamical regime are evaluated. These are the  recurrence time statistics and the diffusion and drift properties of system. An intimate connection is observed between the dynamical 
and statistical properties of the system.

The predominant dynamical regimes in the system are the periodic regime, the chaotic structure regime and the mixing regime. These, together with the nature of the inertial particles, viz. whether aerosol or bubble, influence the diffusion, drift and recurrence properties of the system. The mixing regimes of the system show normal diffusion and exponential decay of recurrence times indicating short term correlations in the system. Chaotic structure regimes 
posses inhomogeneous sticky regimes in the phase space, and these contribute power-law tails to the recurrence time distributions and jump length distributions. Superdiffusive behaviour is seen in the periodic regimes of both aerosols and bubbles. However, some trapping regimes are seen in both the aerosol and bubble cases, and these show sub-diffusive behaviour and low drifts, including zero drifts in the case of stationary states.

Our results may have implications in the context of realistic applications. It was found that the average drift is close to zero in the trapping regions.  This may have important consequences in practical contexts. For example, consider flows  with reacting impurities. Chemical species that have low diffusion and low drift rates can get localized and concentrated in certain regions. These can then react with other chemical species and cause their enhancement or depletion, with further consequences for the environment.

Rapid transport, as in the case of superdiffusive case, can  provide a mechanism by which impurities can easily access  wider regions in the available space. This maybe desirable or undesirable, depending on the context. 
For example smoke emanating from a chimney containing particulate matter with fast diffusion and slow drift can spread the pollutants in the proximity of the source rather than carrying it off to farther places. On the other hand, 
this regime may be highly useful in the context of nutrients spreading through a fluid.

The phase diagrams of the system indicate that a rich variety of dynamical and statistical regimes are available to inertial particles in fluids. Specific choices of regime may be suitable for a specific application. Hence the insights gained from these diagrams may be useful in varied application contexts. 

\begin{acknowledgements}
N. N. T thanks CSIR, India for financial support and N. G thanks DST, India for partial financial support under the project No. SP/S2/HEP/10/2003.
\end{acknowledgements}

\end{document}